\documentclass[useAMS,usenatbib]{mn2e}
\usepackage{acronym} \usepackage{longtable} \usepackage{graphicx}
\usepackage{amsmath} \usepackage{multirow} \usepackage{lscape}
\usepackage{times} \usepackage{amssymb} \usepackage{url}
\usepackage[colorinlistoftodos]{todonotes} \usepackage{color,soul}
\usepackage{array}

\acrodef{ghasp}[\textsc{ghasp}]{Gassendi H-Alpha survey of Spirals}
\acrodef{sdss}[\textsc{sdss}]{Sloan Digital Sky Survey}
\acrodef{ohp}[\textsc{ohp}]{Observatoire de Haute-Provence}
\acrodef{ned}[\textsc{ned}]{NASA/IPAC Extragalactic Database}

\newcommand{\Rc}{R$_{\rm c}$}

\newcommand{\aj}{AJ} \newcommand{\aaa}{A\&A} \newcommand{\aap}{A\&A}
\newcommand{\aas}{A\&AS} \newcommand{\aaps}{A\&AS} \newcommand{\apj}{ApJ}
\newcommand{\apjl}{ApJL} \newcommand{\apjs}{ApJS} \newcommand{\mnras}{MNRAS}

\newcommand{\Deg}{${}^{\circ}$} \newcommand{\araa}{ARA\&A}

\title[GHASP X: Rc-band Photometry]{GHASP: an H$\alpha$ kinematic survey of
spiral galaxies - X. Surface photometry, decompositions and the Tully-Fisher
relation in the \Rc-band\thanks{Based on observations performed at Observatoire
de Haute Provence, France}} \author[Barbosa et al.]{ C. E.
Barbosa$^{1}$\thanks{Corresponding author: carlos.barbosa@usp.br}, C. {Mendes de
Oliveira}$^{1}$, P. Amram$^{2}$, F. Ferrari$^{3}$, \newauthor D. Russeil$^{2}$,
B. Epinat$^{2}$, V. Perret$^{2}$, C. Adami$^{2}$, M. Marcelin$^{2}$\\
$^{1}$Universidade de S\~{a}o Paulo, IAG, Departamento de Astronomia, 
Rua do Mat\~{a}o 1226, Cidade Universit\'aria, 05508-090, S\~{a}o Paulo, SP,
Brazil\\ $^{2}$Aix Marseille Universit\'e, CNRS, LAM (Laboratoire
d'Astrophysique de Marseille), 13388, Marseille, France\\ $^{3}$Instituto de
Matem\'{a}tica, Estat\'{i}stica e F\'{i}sica, FURG, 96201-900, Rio Grande, RS,
Brazil\\ }

\begin{document}

\date{Accepted 2015 July 22.  Received 2015 July 6; in original form 2015 February 22}

\pagerange{\pageref{firstpage}--\pageref{lastpage}} \pubyear{2014}

\maketitle

\label{firstpage}

\begin{abstract}

We present \Rc-band surface photometry for 170 of the 203 galaxies in GHASP, Gassendi H-Alpha survey of SPirals, a sample of late-type galaxies for which high-resolution Fabry-Perot Hα maps have previously been obtained. Our data set is constructed by new \Rc-band observations taken at the Observatoire de Haute-Provence (OHP), supplemented
with Sloan Digital Sky Survey (SDSS) archival data, obtained with the purpose of deriving homogeneous photometric profiles and parameters. Our results include \Rc-band surface brightness profiles for 170 galaxies and $ugriz$ profiles for 108 of these objects. We catalogue several parameters of general interest for further reference, such as total magnitude, effective radius and isophotal parameters -- magnitude, position angle, ellipticity and inclination. We also perform a structural decomposition of the surface brightness profiles using a multi-component method in order to separate disks from bulges and bars, and to observe the main scaling relations involving luminosities, sizes and maximum velocities. 

We determine the \Rc-band Tully Fisher relation using maximum velocities derived solely from H$\alpha$ rotation curves for a sample of 80 galaxies, resulting in a slope of $-8.1 \pm 0.5$, zero point of $-3.0 \pm 1.0$ and an estimated intrinsic scatter of $0.28 \pm 0.07$. We note that, different from the TF-relation in the near-infrared derived for the same sample, no change in the slope of the relation is seen at the low-mass end (for galaxies with $V_{max} < 125$ km/s). We suggest that this different behaviour of the Tully Fisher relation (with the optical relation being described by a single power-law while the near-infrared by two) may be caused by differences in the stellar mass to light ratio for galaxies with $V_{max} < 125$ km/s. 
\end{abstract}


\section{Introduction}

\acresetall

Historically, spiral galaxies have performed a critical role in the studies of
the dark matter. Observations of the outer flat rotation curves in spiral
galaxies \citep[e.g.][]{1978ApJ...225L.107R} have focused the attention to the
then overlooked missing mass problem \citep[see, e.g.][]{1937ApJ....86..217Z},
that stresses the fact that most of what we see (light) is just a fraction of what
we would like to observe (mass). A critical further step, yet to be accomplished, is
to understand the connection between ordinary and dark matter in the inner
regions of galaxies to understand whether (and possibly how) light traces mass. 

The kinematic decomposition of velocity fields of spiral galaxies is the general
method to map their distribution of dark matter
\citep[e.g.,][]{1985ApJ...295..305V,1986RSPTA.320..447V,1986AJ.....91.1301K,2006ApJ...643..804K}.
However, the stellar mass distribution is poorly constrained, and the
under-determined stellar mass-to-light ratio (M/L) translates into degeneracies,
such as the disk-halo and the cusp-core problems, that prevent unique
decompositions. In this context, high resolution, accurate rotation curves, such
as the observed by  the \ac{ghasp}, are necessary to alleviate the problem
\citep{2005ApJ...619..218D}.  

Previous works have supported the scenario of cored dark matter profiles
\citep[e.g.][]{2008MNRAS.383..297S}, but studies on the systematic errors and
larger, homogeneous samples, are still needed to confirm these results. This
series of papers on the \ac{ghasp} survey has the goal of imposing tighter
constraints on the study of dark matter distributions in spiral galaxies. In
this paper, we build a new surface photometry data set for 128 \ac{ghasp}
galaxies in the \Rc-band, observed over the years at the \ac{ohp}, which
provides the basis for the determination of stellar masse in forthcoming work.
Additionally, we complement this data with public \ac{sdss} data in order to
obtain $ugriz$ photometry for 108 \ac{ghasp} galaxies as well as to increase the
\Rc-band data to 170 galaxies ($\approx 84\%$ of the survey). 

Besides the surface brightness profiles, we also compile a homogeneous
photometric catalogue including several photometric quantities of general
interest, such as magnitudes, sizes and isophotal properties. In addition, we
perform a multi component light decomposition in order to separate the light
from the disks (our main interest to dynamical decomposition) from other
components such as bulges and bars. Finally, we perform a first set of
applications to our data set by determining important scaling relations with
luminosity, size and velocity of galaxies, and by deriving the Tully
Fisher relation in the \Rc-band. 

This paper is organized as follows. The \ac{ghasp} sample is briefly outlined in
Section  \ref{sec:sample}. Following this, the details of the observations, data
reduction and calibration are shown in Section \ref{sec:observations}. In
Section \ref{sec:analysis} we present the methods used for the determination of
the surface brightness, PA, ellipticity and integrated magnitude profiles, and
we detail the process of multi component decomposition. In Section
\ref{sec:tests}, we test our results against other similar works, and we check
the internal consistency of our results. Finally, in Section
\ref{sec:scaling} we derive several scaling relations involving luminosity, size and
rotation velocity using the decomposition results, with special emphasis on the
\Rc-band Tully-Fisher relation. 

\section{The GHASP sample} \label{sec:sample} The \textsc{ghasp} sample consists
of 203 spiral and irregular galaxies in the local universe for which
high-resolution H$\alpha$ maps have been observed with Fabry-Perot
interferometry
\citep{2002A&A...387..821G,2003A&A...399...51G,2004MNRAS.349..225G,2005MNRAS.362..127G,2008MNRAS.383..297S,2008MNRAS.388..500E,2008MNRAS.390..466E,2010MNRAS.401.2113E,2011MNRAS.416.1936T}.
The \ac{ghasp} sample was initially designed to be a subsample of the Westerbork
survey \citep[\textsc{WHISP},][]{2001ASPC..240..451V} with the goal of providing
a local universe reference for kinematics and dynamics of disk-like galaxies. 

\begin{figure} \centering \includegraphics[width=\linewidth]{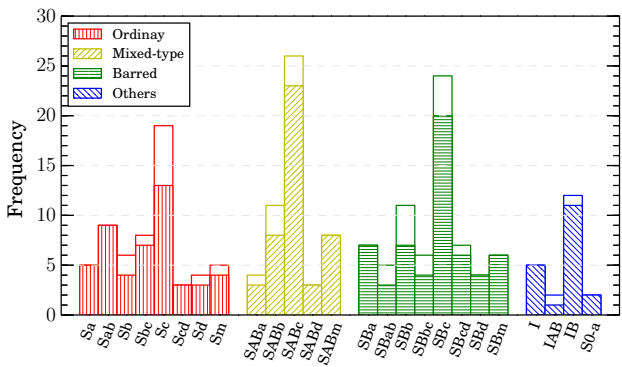}
\caption{Frequency of galaxy morphological types of the \ac{ghasp} sample
according to the Hyperleda classification, separated in sections considering
only ordinary spirals (59 galaxies), mixed types (52), barred (70), irregulars
(19) and lenticulars (2).} \label{fig:morphs} \end{figure}

The \ac{ghasp} sample was designed to cover a large range of morphological
types, including ordinary, mixed-type and barred galaxies, thus excluding only
early-type galaxies because of their low H$\alpha$ content, as illustrated in
Figure \ref{fig:morphs}. The photometric sample presented here is built with
data coming from two sources. Photometric \Rc-band data was obtained by the
\textsc{ghasp} collaboration at the \ac{ohp} over the last decade for 128
galaxies. To enlarge the sample, we also take advantage of the public dataset
from the seventh Data Release (DR7) of the \ac{sdss}
\citep{2002AJ....123..485S}, which provides imaging and calibration in five pass
bands ($ugriz$) for 108 of our galaxies. By the combination of both data sets,
we are able to obtain photometry of a total of 170 \ac{ghasp} galaxies, which
are listed in table \ref{tab:sample} with details about the observation.
However, we note that, on average, the data observed at OHP  in the \Rc-band
goes about half magnitude deeper than the SDSS data. 

\begin{table*} \caption{Photometric sample of the GHASP survey. The printed
version contains only an abridged version of the table, with the remaining
material available online as supplementary material. (1) Galaxy name. (2-3)
Right ascension and declination of the galaxies according to the NED database
(4-5) Morphological classification according to the Hubble type and to the de
Vaucouleurs numerical type from the Hyperleda database. (6) Distance to the
galaxies according to \citet{2008MNRAS.388..500E}. (7) Shows if the galaxy is in
the SDSS. (8-10) OHP \Rc-band observation log, including the runs, total
exposure time and the seeing.} \begin{tabular}{cccccccccc} \hline \hline &
&          &            &          &      &  &
\multicolumn{3}{c}{\underline{\hspace{.5cm}OHP observation log.\hspace{.5cm}}}\\
Galaxy & $\alpha$ & $\delta$ & Morphology & Morphology & Distance & SDSS & Runs
& Exptime & FWHM \\ & (J2000)  & (J2000)  &            &  t & (Mpc)   &      &
& (s)     & (arcsec) \\
 
(1) & (2) & (3) & (4) & (5) & (6) & (7) & (8) & (9) & (10)\\ \hline UGC 89 &
00h09m53.4s & +25d55m26s & SBa & $ 1.2\pm 0.6$ & 64.2 & no & 2,5 & 600 & 2.7\\
UGC 94 & 00h10m25.9s & +25d49m55s & S(r)ab & $ 2.4\pm 0.6$ & 64.2 & no & 5,6 &
3300 & 2.3\\ IC 476 & 07h47m16.3s & +26d57m03s & SABb & $ 4.2\pm 2.6$ & 63.9 &
yes & --- & --- & ---\\ UGC 508 & 00h49m47.8s & +32d16m40s & SBab & $ 1.5\pm
0.9$ & 63.8 & no & 5 & 3600 & 2.3\\ UGC 528 & 00h52m04.3s & +47d33m02s & SABb &
$ 2.9\pm 1.1$ & 12.1 & no & 2 & 1500 & 1.9\\ NGC 542 & 01h26m30.9s & +34d40m31s
& Sb pec & $ 2.8\pm 3.9$ & 63.7 & yes & --- & --- & ---\\ UGC 763 & 01h12m55.7s
& +00d58m54s & SABm & $ 8.6\pm 1.0$ & 12.7 & yes & 2,5 & 600 & 3.1\\ UGC 1013 &
01h26m21.8s & +34d42m11s & SB(r)b pec & $ 3.1\pm 0.2$ & 70.8 & yes & 2,5 & 5100
& 2.5\\ UGC 1117 & 01h33m50.9s & +30d39m37s & Sc & $ 6.0\pm 0.4$ & 0.9 & no & 5
& 4500 & 2.7\\ UGC 1249 & 01h47m29.9s & +27d20m00s & SBm pec & $ 8.8\pm 0.6$ &
7.2 & no & 2 & 1800 & 2.1\\ ...      & ...         &   ...      & ...     & ...
& ... & ... &  ...  & ... & ... \\ UGC 11951 & 22h12m30.1s & +45d19m42s & SBa &
$ 1.1\pm 0.8$ & 17.4 & no & 2 & 2100 & 2.6\\ UGC 12060 & 22h30m34.0s &
+33d49m11s & IB & $ 9.9\pm 0.5$ & 15.7 & no & 6 & 6000 & 2.5\\ UGC 12082 &
22h34m10.8s & +32d51m38s & SABm & $ 8.7\pm 0.8$ & 10.1 & no & 6 & 3000 & 3.7\\
UGC 12101 & 22h36m03.4s & +33d56m53s & Scd & $ 6.6\pm 0.9$ & 15.1 & yes & 2 &
1800 & 1.9\\ UGC 12212 & 22h50m30.3s & +29d08m18s & Sm & $ 8.7\pm 0.5$ & 15.5 &
yes & 2 & 1800 & 2.2\\ UGC 12276 & 22h58m32.5s & +35d48m09s & SB(r)a & $ 1.1\pm
0.5$ & 77.8 & no & 2 & 2700 & 2.0\\ UGC 12276c & 22h58m32.5s & +35d48m09s & S? &
$ 5.1\pm 5.0$ & 77.8 & no & 2 & 2700 & 2.0\\ UGC 12343 & 23h04m56.7s &
+12d19m22s & SBbc & $ 4.4\pm 0.9$ & 26.9 & no & 2,5 & 1500 & 2.7\\ UGC 12632 &
23h29m58.7s & +40d59m25s & SABm & $ 8.7\pm 0.5$ & 8.0 & no & 5 & 4500 & 2.4\\
UGC 12754 & 23h43m54.4s & +26d04m32s & SBc & $ 6.0\pm 0.4$ & 8.9 & no & 2 & 1200
& 2.3\\ \hline \hline \end{tabular} \label{tab:sample} \end{table*}

\section{Data reduction} \label{sec:observations}

\subsection{Data from the \acs{ohp} observatory} \label{sec:dataohp}

Broadband imaging for 128 galaxies in the \Rc-band was obtained with the 1.2m
telescope at the \acf{ohp}, France, in several observation runs as presented in
Table \ref{tab:obsruns}. The images have a field of view of 11.7' x 11.7', taken
with a single CCD with 1024 x 1024 pixels, resulting in a pixel size of 0.685
arcsec$^{-1}$. 

\begin{table} \centering \caption{List of observational runs in which the \Rc
photometry of the \textsc{ghasp} galaxies was obtained at \textsc{ohp}.}
\label{tab:obsruns} \begin{tabular}{clc} \hline Run&Dates&Number of\\ &
&Galaxies observed\\ \hline (1) & 2002 Mar 7th - Mar 13rd& 29\\ (2) & 2002 Oct
28th - Nov 10th& 38\\ (3) & 2003 Mar 8th - Mar 9th& 17\\ (4) & 2003 Mar 29th -
Apr 6th& 22\\ (5) & 2003 Sep 22nd - Sep 28th& 25\\ (6) & 2003 Oct 21st - Oct
25th& 6\\ (7) & 2008 Jun 2nd - Jun 4th& 11\\ (8) & 2009 Oct 23rd& 1\\ (9) & 2010
Mar 19th - Mar 21st& 16\\ \hline \end{tabular} \end{table}

Basic data reduction was performed with \textsc{iraf}\footnote{\textsc{Iraf} is
distributed by the National Optical Astronomy Observatory, which is operated by
the Association of Universities for Research in Astronomy (AURA) under
cooperative agreement with the National Science Foundation.} tasks, including
flat-field, bias subtraction and cosmic-ray cleaning. Images of the same galaxy
are then aligned and combined for the cases with roughly the same smallest
seeing FWHM, estimated from isolated field stars. Photometric stability check
and zero point calibration was obtained by the observation of several standard
stars from the catalog of \citet{1992AJ....104..340L} in different times during
the nights, considering the mean airmass correction coefficient of 0.145 for the
\Rc-band \citep{1991A&AS...90..225C}, and no colour term. 

The determination of the sky level is the greatest source of uncertainty for
surface brightness profiles and magnitudes \citep{1996ApJS..103..363C}. For this
purpose, we adopt the method of estimating the background by selecting sky
``boxes'' on the images, which are visually selected areas where the galaxy and
stellar light contribution is minimal, and use those regions to calculate a
smooth surface using the  \textsc{iraf} package \textsc{imsurfit} with
polynomials of order 2,  which is subtracted from the original images. This
process resulted in an homogeneous background for which the typical residual
standard deviation is in the range 0.5-1\% of the sky level. 

We have modeled the Point Spread Function (PSF) of our images using the
\textsc{iraf} \textsc{psfmeasure} task. We selected bright, unsaturated stars
across the fields using the task \textsc{daofind}, and then  modeled their light
profiles using a circular Moffat function \citep{1969A&A.....3..455M}, given by

\begin{equation} \text{PSF}(r) = \frac{\beta-1}{\pi \alpha^2}\left [ 1 + \left (
\frac{r}{\alpha}\right )^2\right ]^{-\beta}\text{,} \end{equation}

\noindent where the radial scale length $\alpha$ and the slope $\beta$ are free
parameters, which can be related to the seeing by the relation
FWHM$=2\alpha\sqrt{2^{1/\beta}-1}$ \citep[see also][]{2001MNRAS.328..977T}. This
method has proved to be suitable in our case due to the presence of extended
wings in the PSFs. Overall, the typical seeing of our observations is
FWHM$\approx$3 arcsec, with the parameters $\alpha$, $\beta$ and FWHM having
mean statistical uncertainties of 1.7\%, 9\% and 3\% respectively. 

\subsection{Data from \acs{sdss}}

To increase the number of galaxies in our photometric sample in the \Rc-band, we
use \ac{sdss} DR7 \citep{2009ApJS..182..543A} archival data for 108
\textsc{ghasp} galaxies we found in the database and transformed \ac{sdss}
$ugriz$ data into \Rc with a multi-band scaling relation (more details in
section \ref{sec:profiles}). Among these 108 galaxies, 66 have also been
observed in the OHP, thus 42 new ones are added to the final photometric sample.
Calibrations and the PSF of the images are obtained directly from the data
products of the survey. We performed a new sky determination for each image for
consistency with the adopted method for \Rc images, and also because a few
authors have pointed out errors in the sky determinations on images with bright
galaxies in early \ac{sdss} releases
\citep[e.g.][]{2007AJ....133.1741B,2007ApJ...662..808L,2007ApJ...660.1186L}.


\section{Data analysis} \label{sec:analysis}

\subsection{Surface photometry} \label{sec:profiles}

We study the photometric properties of the sample using the traditional method
of elliptical isophote fitting \citep{1984ApJS...56..105K,1987MNRAS.226..747J}.
Surface brightness (SB) profiles of the galaxies were obtained using the
\textsc{iraf} task \textsc{ellipse}, which provides a number of parameters that
describe the light of the galaxy as a function of the semi-major axis (which we
simply refer to as the radius, $r$), including the ellipticity ($\varepsilon$),
position angle (PA) and the curve of growth, which quantifies the total apparent
magnitude inside each isophote.

Masks for foreground and background objects were produced interactively in two
steps. Firstly, most objects in the images were detected and masked out with
\textsc{SExtractor} \citep{1996A&AS..117..393B}. Other important sources not
detected by the program, such as saturated stars and stars/galaxies superposed
to the galaxies of interest were then masked during \textsc{ellipse} runs.
Finally, we checked the results inspecting the residual image produced by
subtracting an interpolated model of the galaxy produced with the task
\textsc{bmodel}. This process was carried out several times for each galaxy
until no bright sources were observed in the resulting subtracted images except
for spiral arms and/or bars of the galaxy not masked on purpose. 

The centre of each galaxy was defined in a first iteration of \textsc{ellipse}
and later was set fixed for all other iterations. The position angle and
ellipticity of the isophotes were usually set free to vary as a function of the
radial distance, as usually done for late-type galaxy photometry
\citep[e.g.][]{2003ApJ...582L..79B,2003ApJ...582..689M,2009MNRAS.394.2022M}, but
in a few cases we were forced to fix the geometric parameters in part or in the
whole galaxy in order to obtain convergence for the photometry. For the
\ac{sdss} data, we adopt this method in the r-band images, due to its relatively
high signal-to-noise ratio in the $ugriz$ system, but we fix the position angles
and ellipticities accordingly to the r-band parameters in the other pass bands
in order to obtain consistent colours. 

Uncertainties for the SB profiles include the isophote determination error given
by \textsc{ellipse}, the photon counting statistics of the detector, and the sky
level subtraction uncertainty, all added in quadrature. 	All profiles are
corrected for the Galactic foreground extinction using the dust reddening maps
of \citet{1998ApJ...500..525S}, assuming a dust model with constant selective
extinction of 3.1, and relative extinction for the different pass bands
according to table 6 of  \citet{1998ApJ...500..525S}. However, we do not
attempt for a correction of the SB profiles for the more uncertain problem of
the galaxies' internal extinction. 

\begin{figure} \centering \includegraphics[width=\linewidth]{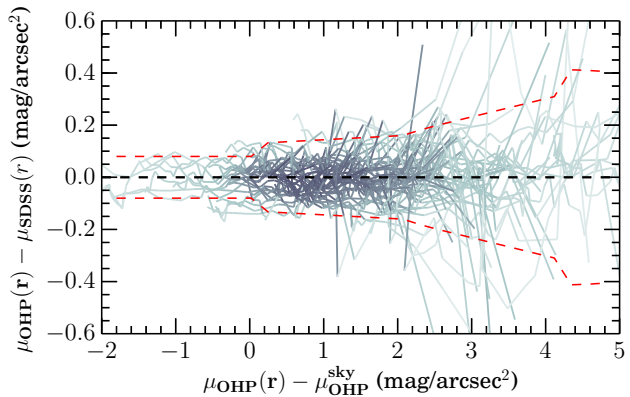}
\caption{Difference between \Rc surface brightness profiles for OHP  and SDSS
data for 54 galaxies in common to both samples as a function of the OHP SB
profile relative to the sky level. Profiles for the \ac{sdss} galaxies are
obtained using equation \eqref{eq:jester}. Each blue line represents the
difference in the surface brightness profiles of a single galaxy profile
comparison. The dashed red lines indicates the running RMS difference between the
profiles.} \label{fig:ugriztoR} \end{figure}

Finally, to obtain \Rc-band SB profiles from \ac{sdss} data, we use a slightly
modified version of the relation derived by \citet{2005AJ....130..873J}, given
by 

\begin{equation} \mu_{R}(r) = 0.42\mu_g(r) - 0.38\mu_r(r) + 0.96\mu_i(r)-0.16
\text{,} \label{eq:jester} \end{equation}

\noindent where $\mu(r)$ represents the surface brightness profile at radius $r$
in the pass band indicated in the subscripts. The equation above was originally
derived for stellar photometry, so we have tested its accuracy in surface
photometry by the comparison of OHP SB profiles, obtained directly in the
\Rc-band, with profiles derived from the SDSS $ugriz$ bands using a sample of 54
galaxies for which the geometric parameters of both data sets are similar. The
results are presented in Figure \ref{fig:ugriztoR}, which shows the difference
of the profiles as a function of the OHP surface brightness profiles relative to
the sky level, which varies in the OHP observations, from one galaxy observation to another. The red line shows the
running RMS difference between the profiles, indicating that the error in the
process of transforming between the photometric systems is of $\sim 0.08$ mag
arcsec$^{\rm{-2}}$ in the regions brighter than the sky level, $\sim 0.15$ mag
arcsec$^{\rm{-2}}$ for the regions down to two magnitudes fainter than the sky,
and $\sim 0.4$ mag arcsec$^{\rm{-2}}$ for the regions 5 magnitudes fainter than
the sky level. 

\begin{figure*} \centering
\includegraphics[width=0.295\textwidth]{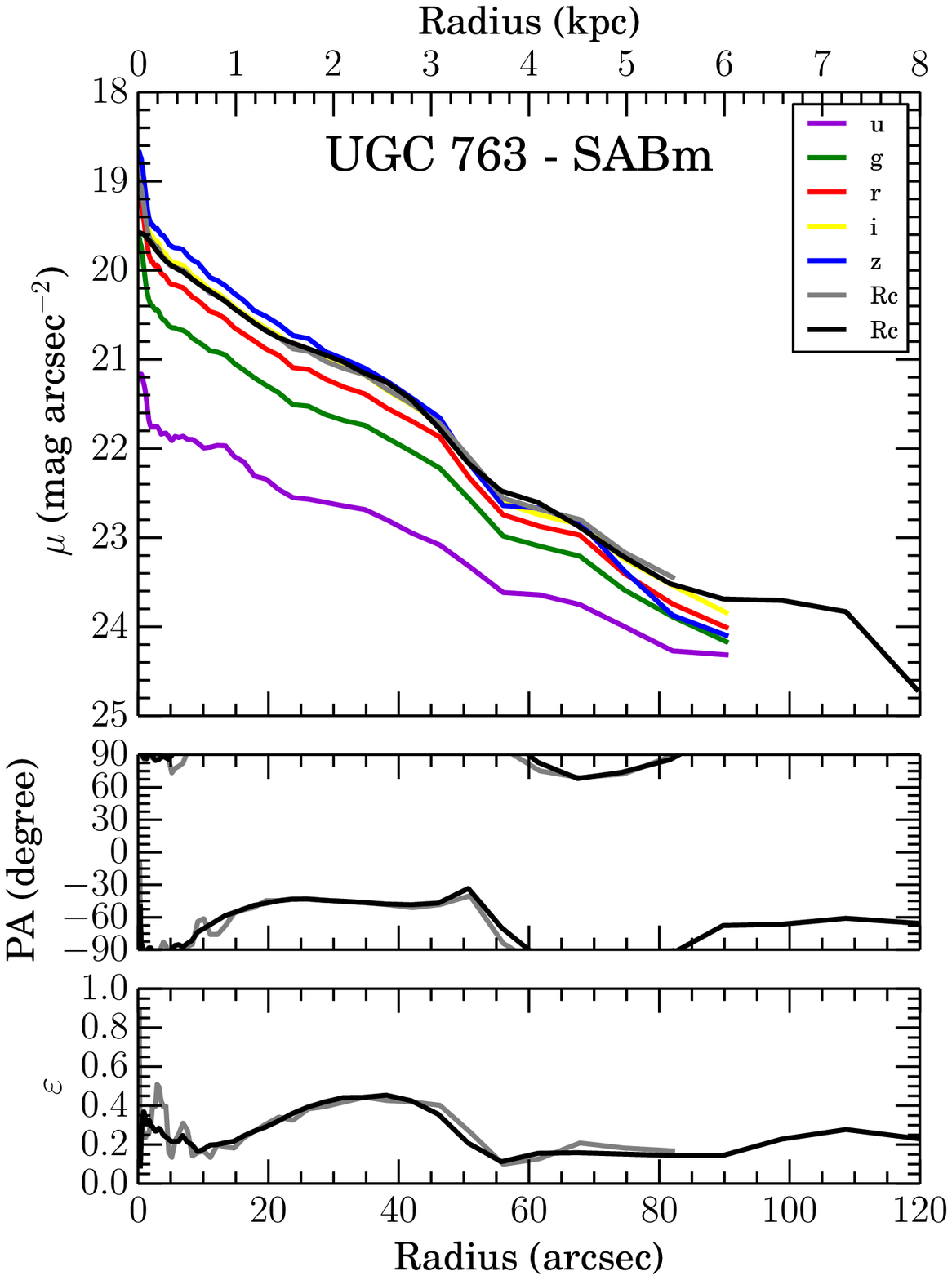}
\includegraphics[width=0.295\textwidth]{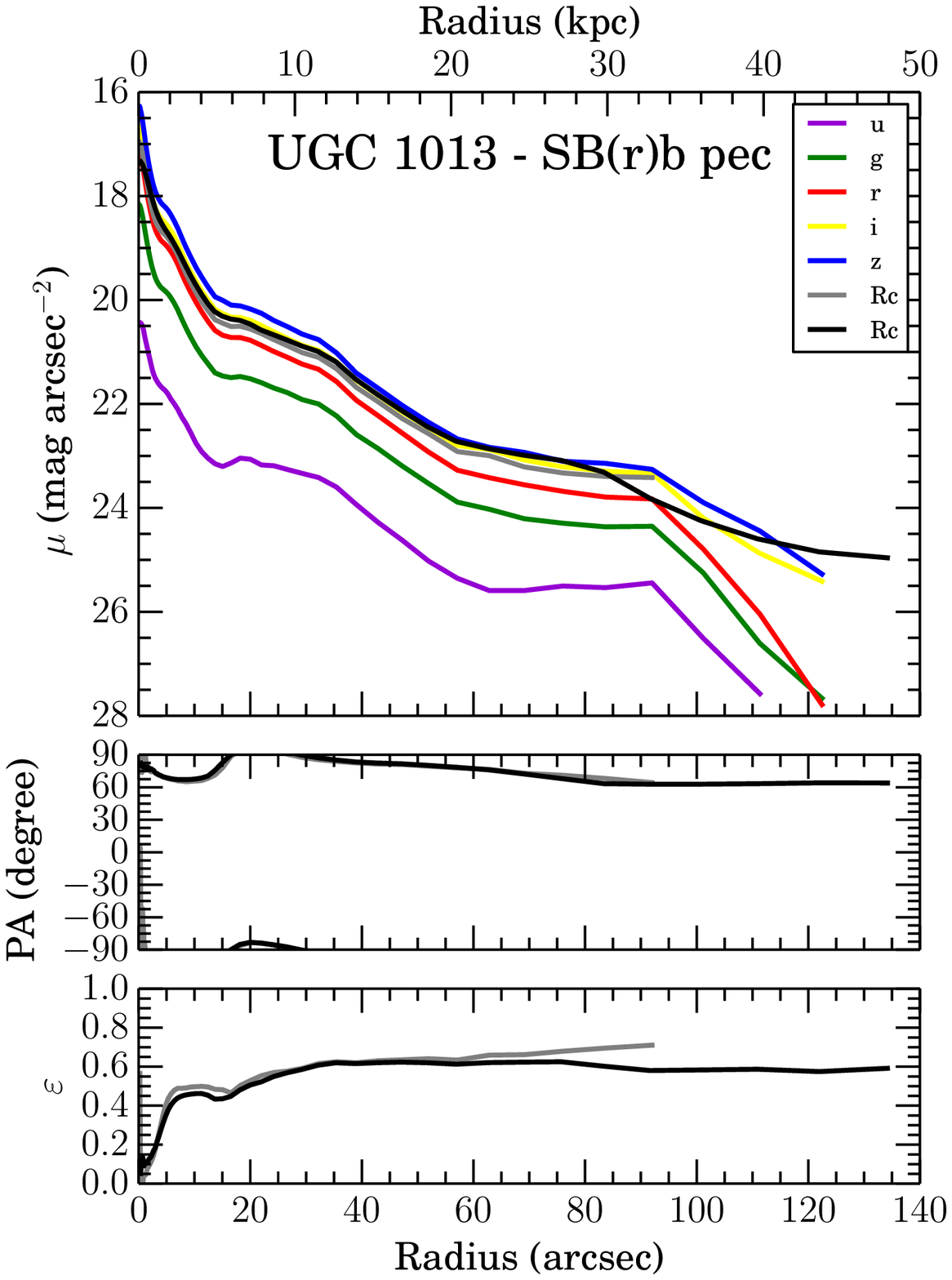}
\includegraphics[width=0.295\textwidth]{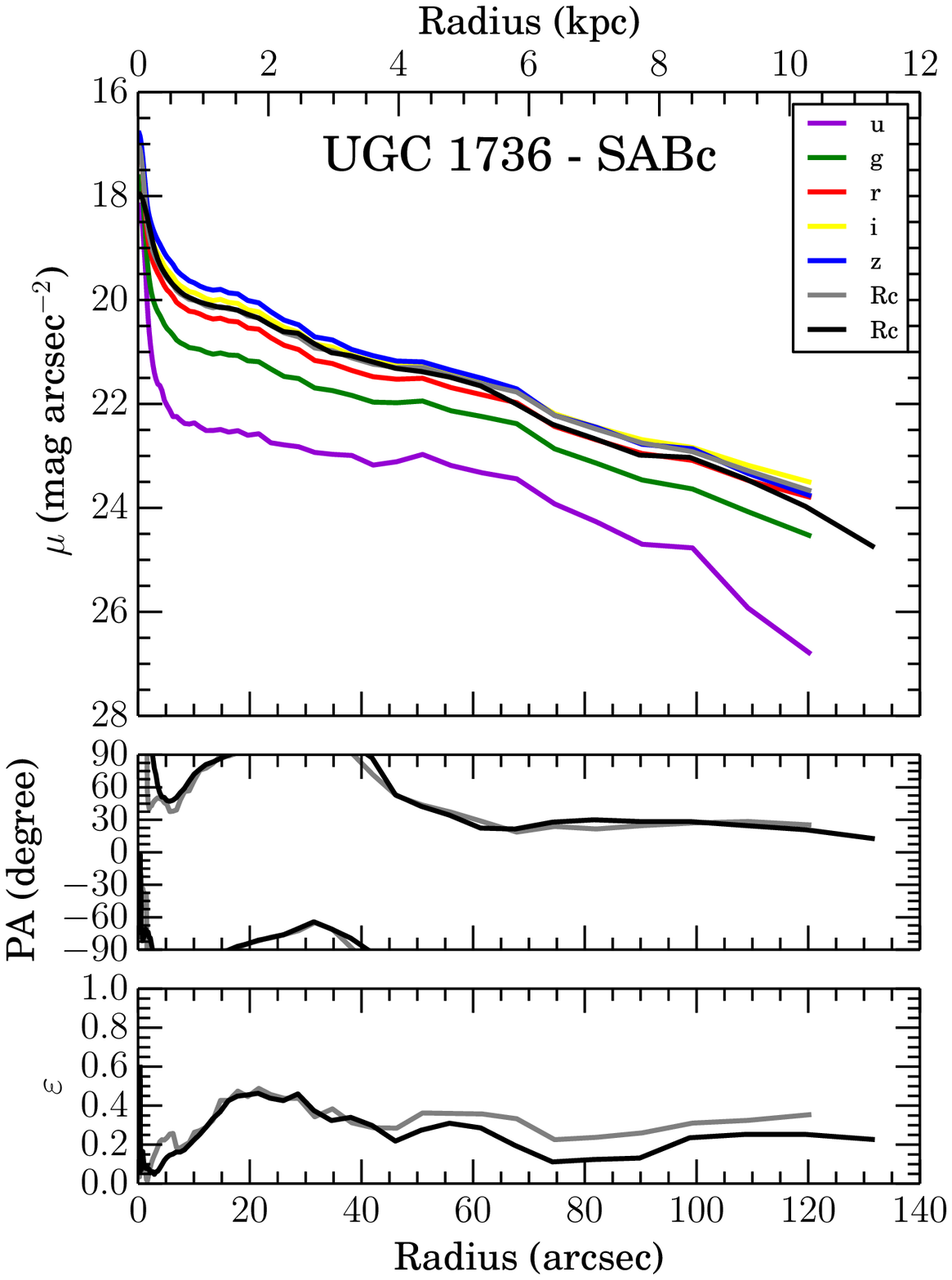}
\includegraphics[width=0.295\textwidth]{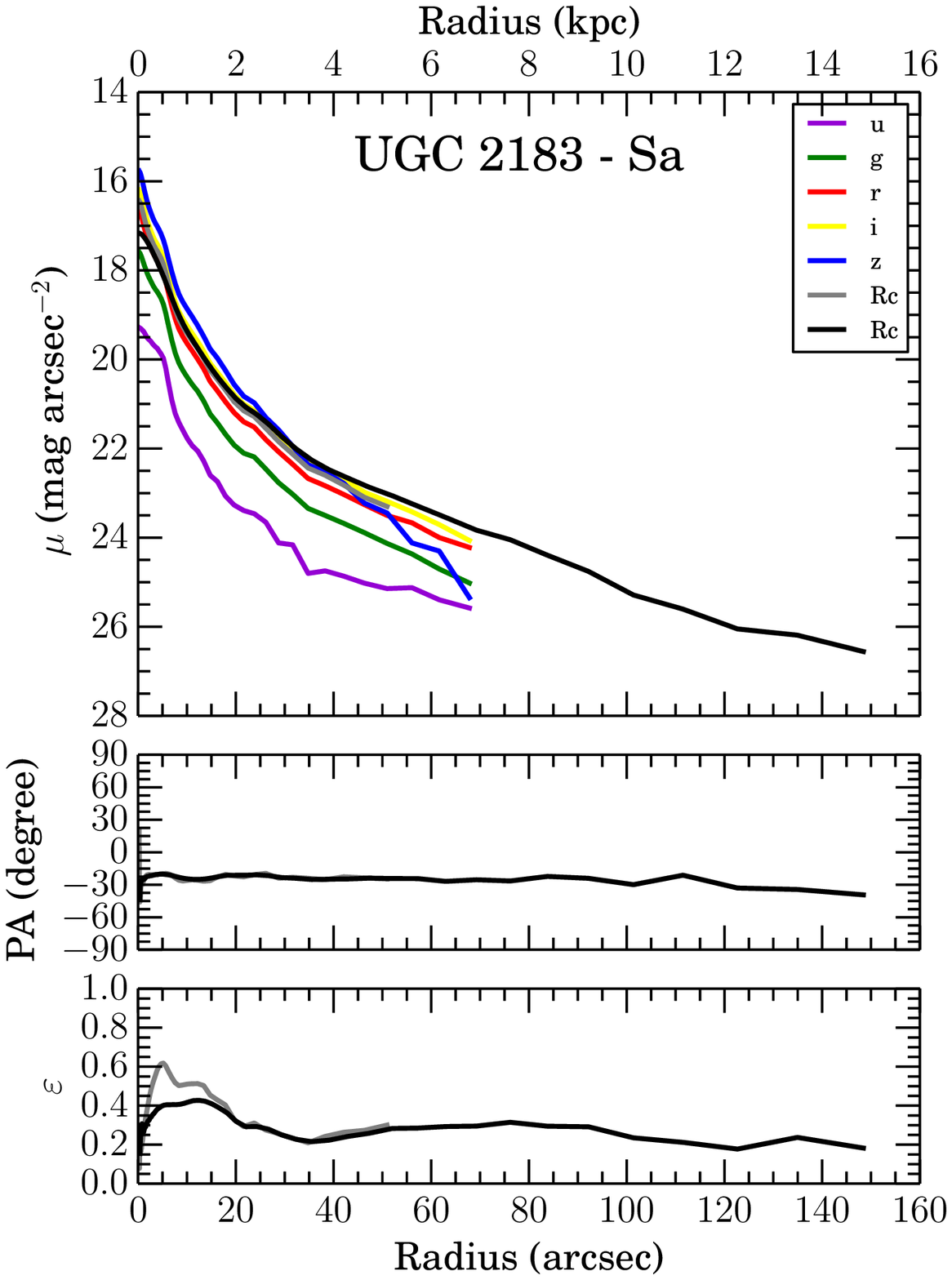}
\includegraphics[width=0.295\textwidth]{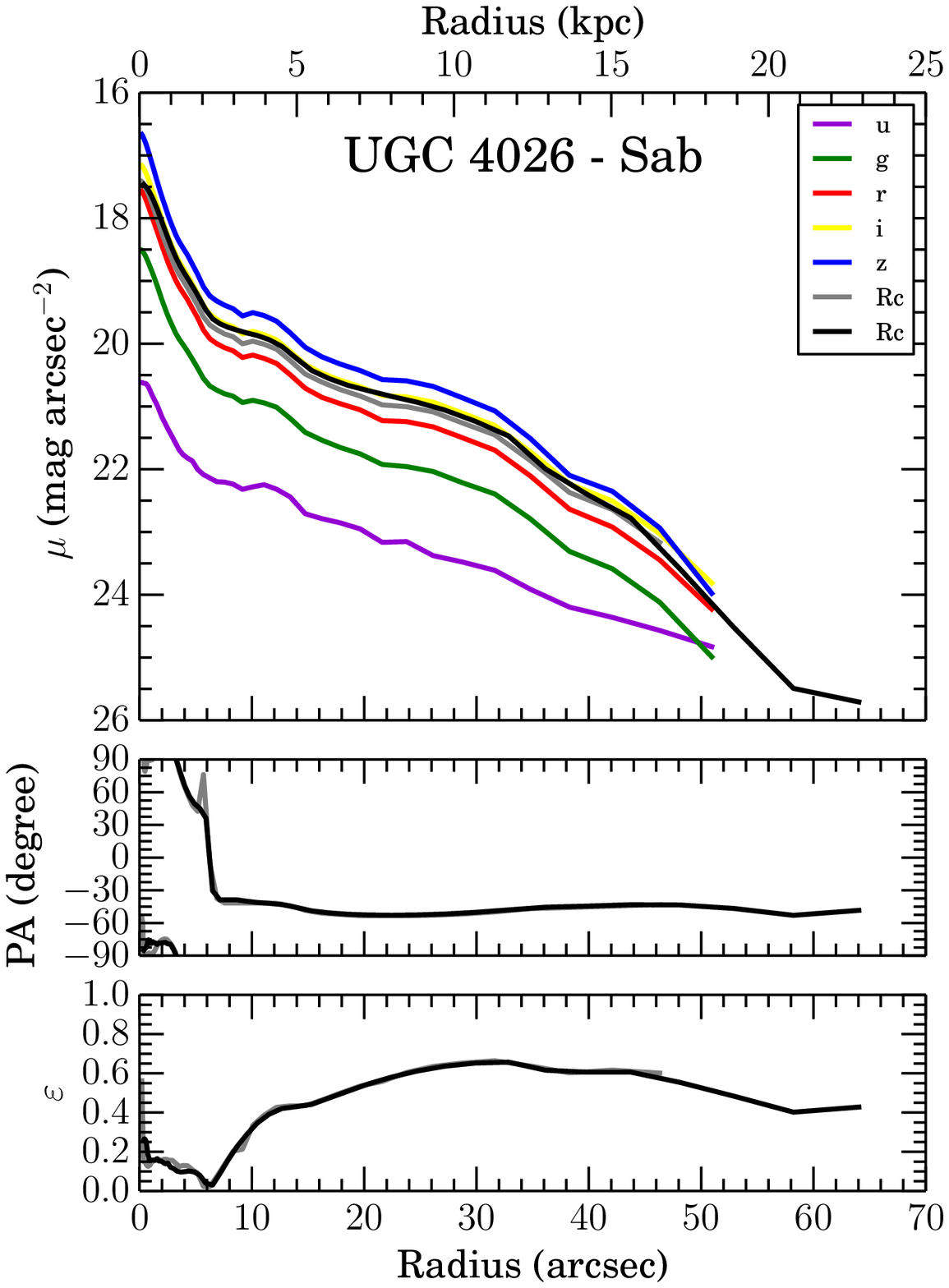}
\includegraphics[width=0.295\textwidth]{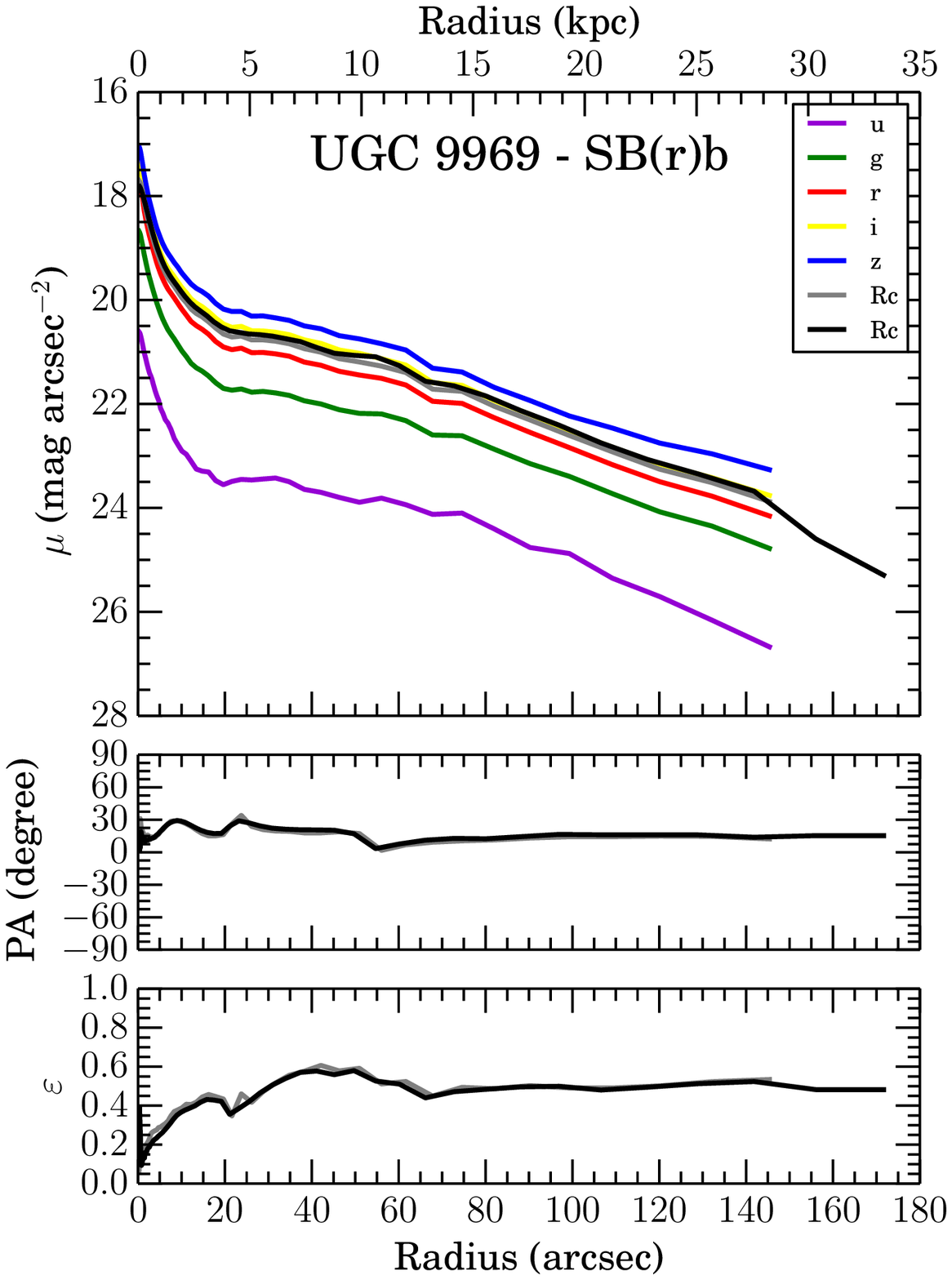} \caption{
Examples of surface brightness profiles in the \Rc-band, position angle and
ellipticity profiles for six galaxies for which data are available both for the
OHP and for the SDSS data sets. The surface brightness profiles are presented in
different colours for each band according to the upper right labels, and
uncertainties are not presented for the sake of clarity. The middle and lower panels of each
galaxy show the position angle and ellipticity profiles respectively, with gray
representing the r-band results and black representing the \Rc-band geometry.
All profiles are corrected for the Galactic foreground extinction according to
the dust maps of \citet{1998ApJ...500..525S}.} \label{fig:sbexamples}
\end{figure*}

We present a sample of SB profiles for a variety of morphological types in
Figure \ref{fig:sbexamples}, including also the ellipticity and position angle
variations. All surface brightness profiles are available in electronic format.
In the next section, we detail other catalogued \Rc-band photometric properties
derived from SB profiles in this section. 

\subsection{Integrated and isophotal photometry in the \Rc-band}
\label{sec:intphot}

For the \Rc-band SB profiles derived in this work, we obtained a
number of properties of the galaxies which are of general interest by fixing a
reference isophotal level. In the case of the \Rc-band, the isophotal level of
23.5 mag arcsec$^{\text{-2}}$ is usually used as reference, because it
corresponds to an aperture similar to the B-band isophote of 25 mag
arcsec$^{\text{-2}}$. However, this level was reached for only 72\% of our
surface brightness profiles. Therefore, in order to provide a more complete
catalogue for our sample, we also use the isophotal level of 22.5 mag
arcsec$^{\text{-2}}$ to provide parameters for 98\% of the sample. 

We measured the isophotal radius ($r_{\rm iso}$), position angle (PA$_{\rm
iso}$), ellipticity ($\varepsilon_{\rm iso}$) and apparent magnitude ($m_{\rm
R,iso}$) directly from the SB profiles, with uncertainties estimated by Monte
Carlo simulations of perturbations of the profile according to their
uncertainties. Also, the inclination of the galaxies is estimated at a given
isophotal level as \citep{1988cng..book.....T} 

\begin{equation} \cos i_{\rm iso} = \frac{(1-\varepsilon_{\rm iso})^2-q_0^2}{1^2
- q_0^2}\mbox{,} \label{eq:inclination} \end{equation}

\noindent where $q_0=0.2$ is the intrinsic flattening of edge-on disks
\citep[e.g.][]{1984AJ.....89..758H,1996ApJS..103..363C}. 

We also measured the total (asymptotic) apparent magnitudes of the galaxies
($m_{\rm R,total}$), which were calculated by the extrapolation of the curve of
growth of the SB profiles using a derivative method similar to
\citet{2001ApJS..136..393C}. However, this method failed in cases of galaxies for
which the curve of growth did not converge. In these cases, we used the
last isophote total magnitude to estimate a lower limit to the total magnitude. Also
using the curve of growth, we measured the effective radius of the galaxies
($r_{50}$), which is defined as the radius containing 50\% of the total light of
the galaxy. In the cases where we have not obtained a safe total magnitude, we then
estimated the lower limits of the effective radius. Uncertainties in these
parameters are also based on Monte Carlo simulations.

Table \ref{table:photR} presents a sample of the results for the isophotal
level of 22.5 mag arcsec$^{\text{-2}}$. The complete catalogue, and the
catalogue for the  isophotal level of 23.5 mag arcsec$^{\text{-2}}$ are provided
in the supplementary material.

\begin{table*} \caption{Isophotal and integrated photometric parameters in \Rc
band. The printed version contains only an abridged version of the table, and
the remaining material is available online. (1) Galaxy name. (2) Data source.
(3-7) Position angle, ellipticity, inclination, radius and apparent integrated
magnitude at the isophote of 22.5 mag arcsec$^{-2}$. (8) Effective radius. (9)
Total apparent magnitude.} \begin{tabular}{ccccccccc} \hline \hline Galaxy &
Data & PA$_{22.5}$ & $\varepsilon_{22.5}$ & $i_{22.5}$ & $r_{22.5}$ & $m_{\rm
R,22.5}$ & $r_{50}$ & $m_{\rm R,total}$ \\ & & (arcsec) & (degree) & (degree) &
(arcsec) & (mag) & (arcsec) & (mag) \\ (1) & (2) & (3) & (4) & (5) & (6) & (7) &
(8) & (9)\\ \hline 
UGC 89 & OHP & $-27\pm11$ & $0.22\pm0.03$ & $40\pm3$ & $41\pm2$ & $11.60\pm0.03$ & $19\pm5$ & $11.4\pm0.1$\\
UGC 94 & OHP & $-84\pm2$ & $0.30\pm0.02$ & $46\pm1$ & $28\pm1$ & $13.12\pm0.03$ & $16\pm7$ & $12.7\pm0.4$\\
IC 476 & SDSS & $-79\pm3$ & $0.4\pm0.1$ & $52\pm9$ & $16\pm2$ & $14.7\pm0.1$ & $>10.0$ & $<14.53$\\
UGC 508 & OHP & $-61\pm4$ & $0.11\pm0.05$ & $28\pm7$ & $67\pm1$ & $11.17\pm0.03$ & $40\pm1$ & $10.91\pm0.08$\\
UGC 528 & OHP & $52\pm2$ & $0.03\pm0.02$ & $13\pm3$ & $65\pm3$ & $10.16\pm0.01$ & $21\pm1$ & $10.10\pm0.04$\\
NGC 542 & SDSS & $-34\pm2$ & $0.73\pm0.01$ & $79.1\pm0.9$ & $23\pm2$ & $14.39\pm0.05$ & $12\pm6$ & $14.2\pm0.6$\\
UGC 763 & OHP & $-75\pm23$ & $0.12\pm0.02$ & $29\pm3$ & $57\pm4$ & $11.61\pm0.06$ & $47\pm5$ & $11.1\pm0.2$\\
UGC 763 & SDSS & $-79\pm35$ & $0.12\pm0.08$ & $29\pm10$ & $56\pm6$ & $11.7\pm0.1$ & $>39.0$ & $<11.46$\\
UGC 1013 & OHP & $80\pm2$ & $0.62\pm0.01$ & $70.6\pm0.7$ & $53\pm1$ & $12.03\pm0.01$ & $27\pm4$ & $11.7\pm0.1$\\
UGC 1013 & SDSS & $80\pm3$ & $0.64\pm0.01$ & $72.2\pm0.8$ & $51\pm5$ & $12.17\pm0.04$ & $23\pm3$ & $11.9\pm0.1$\\
... & ... & ... & ... & ... \\ 
UGC 12082 & OHP & $18\pm2$ & $0.42\pm0.01$ & $56.1\pm0.9$ & $19\pm3$ & $15.0\pm0.2$ & $49\pm23$ & $12.8\pm0.5$\\
UGC 12101 & OHP & $-50\pm2$ & $0.53\pm0.01$ & $64.2\pm0.8$ & $51\pm2$ & $12.50\pm0.03$ & $29\pm2$ & $12.32\pm0.09$\\
UGC 12101 & SDSS & $-50\pm3$ & $0.52\pm0.02$ & $64\pm1$ & $49\pm4$ & $12.66\pm0.06$ & $28\pm21$ & $12.3\pm0.9$\\
UGC 12212 & OHP & $-80\pm5$ & $0.36\pm0.02$ & $52\pm1$ & $14\pm1$ & $15.8\pm0.2$ & $27\pm26$ & $14\pm2$\\
UGC 12212 & SDSS & $-71\pm2$ & $0.3\pm0.01$ & $46.8\pm0.8$ & $12\pm5$ & $15.9\pm0.6$ & $>16.0$ & $<14.77$\\
UGC 12276 & OHP & $-47\pm6$ & $0.23\pm0.01$ & $40.8\pm0.9$ & $30\pm1$ & $13.00\pm0.03$ & $19\pm5$ & $12.6\pm0.2$\\
UGC 12276c & OHP & $82\pm2$ & $0.45\pm0.01$ & $58.3\pm0.7$ & $7\pm1$ & $17.0\pm0.2$ & $>5.0$ & $<16.62$\\
UGC 12343 & OHP & $38\pm2$ & $0.30\pm0.03$ & $47\pm2$ & $102\pm1$ & $10.33\pm0.03$ & $71\pm9$ & $10.2\pm0.2$\\
UGC 12632 & OHP & $22\pm2$ & $0.53\pm0.05$ & $64\pm4$ & $24\pm5$ & $14.8\pm0.3$ & $66\pm59$ & $13\pm1$\\
UGC 12754 & OHP & $-13\pm3$ & $0.29\pm0.01$ & $46\pm1$ & $89\pm3$ & $11.16\pm0.03$ & $47\pm2$ & $10.9\pm0.1$\\
\hline \hline \end{tabular} \label{table:photR} \end{table*}

\subsection{Multicomponent Decomposition} \label{sec:decomp}

In our forthcoming work (Pineda et al. in preparation), we plan to study the
kinematic properties of a subsample of \ac{ghasp} galaxies with specific
photometric properties depending on the relative importance of the disks in
comparison with bulges and bars. In order to separate the SB brightness profiles
into different structural components, we proceed to a multicomponent
decomposition of the SB profiles.  

For that purpose, we use a parametric profile fitting method which includes as
many components as necessary to separate the photometric components -- disks,
bulges, bars, spiral arms, lens and nuclear sources. Ideally, one could use a
complete 2D fitting to better describe the non axisymmetric components, like the
bar, but the reliability of the 1D method to recover the structural parameters
is comparable to 2D, at least for the disk component
\citep{2003ApJ...582..689M}, and also allows the estimation of the integrated
properties with good accuracy. 

\begin{figure*} \centering
\includegraphics[width=0.49\textwidth]{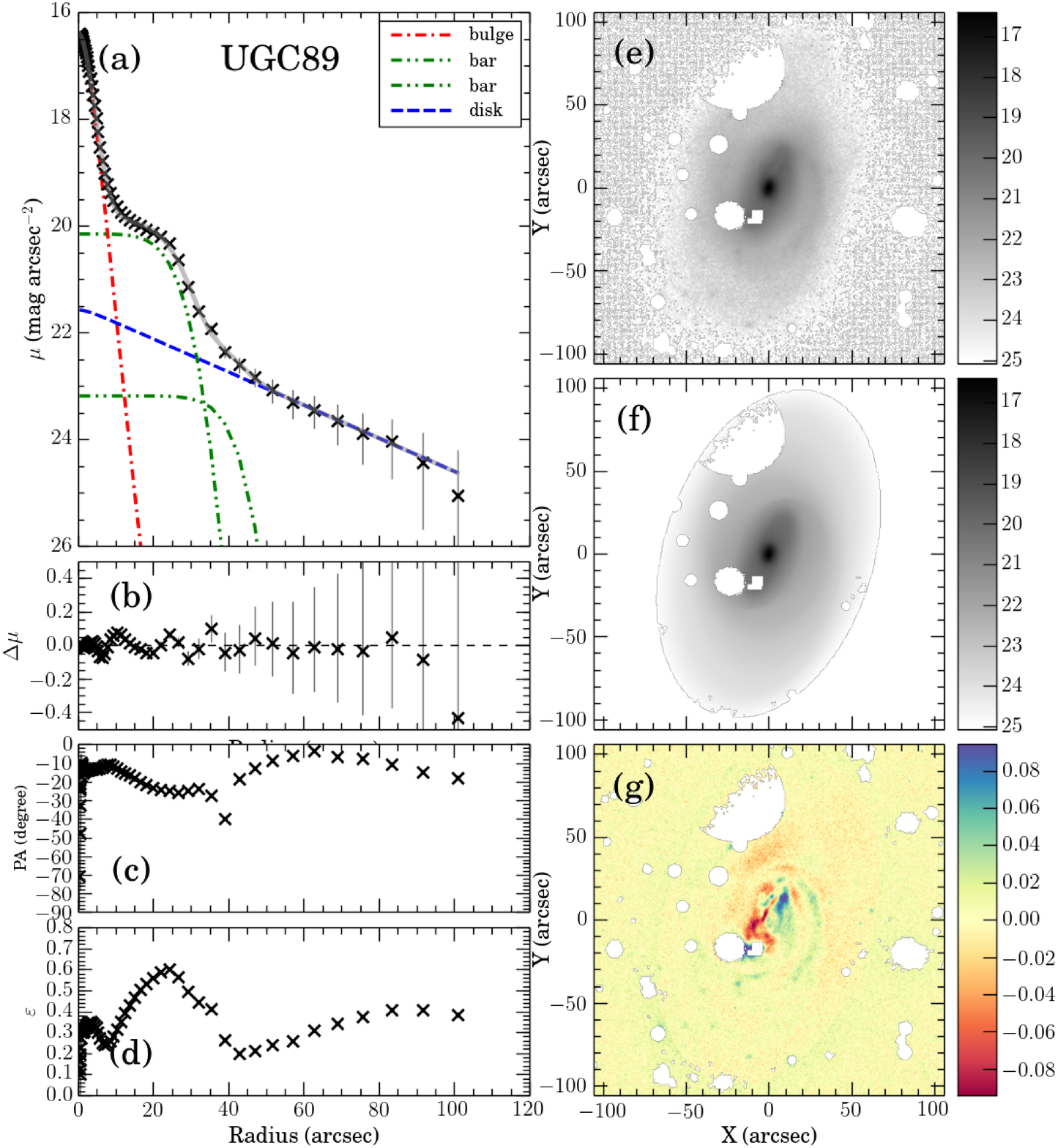}
\includegraphics[width=0.49\textwidth]{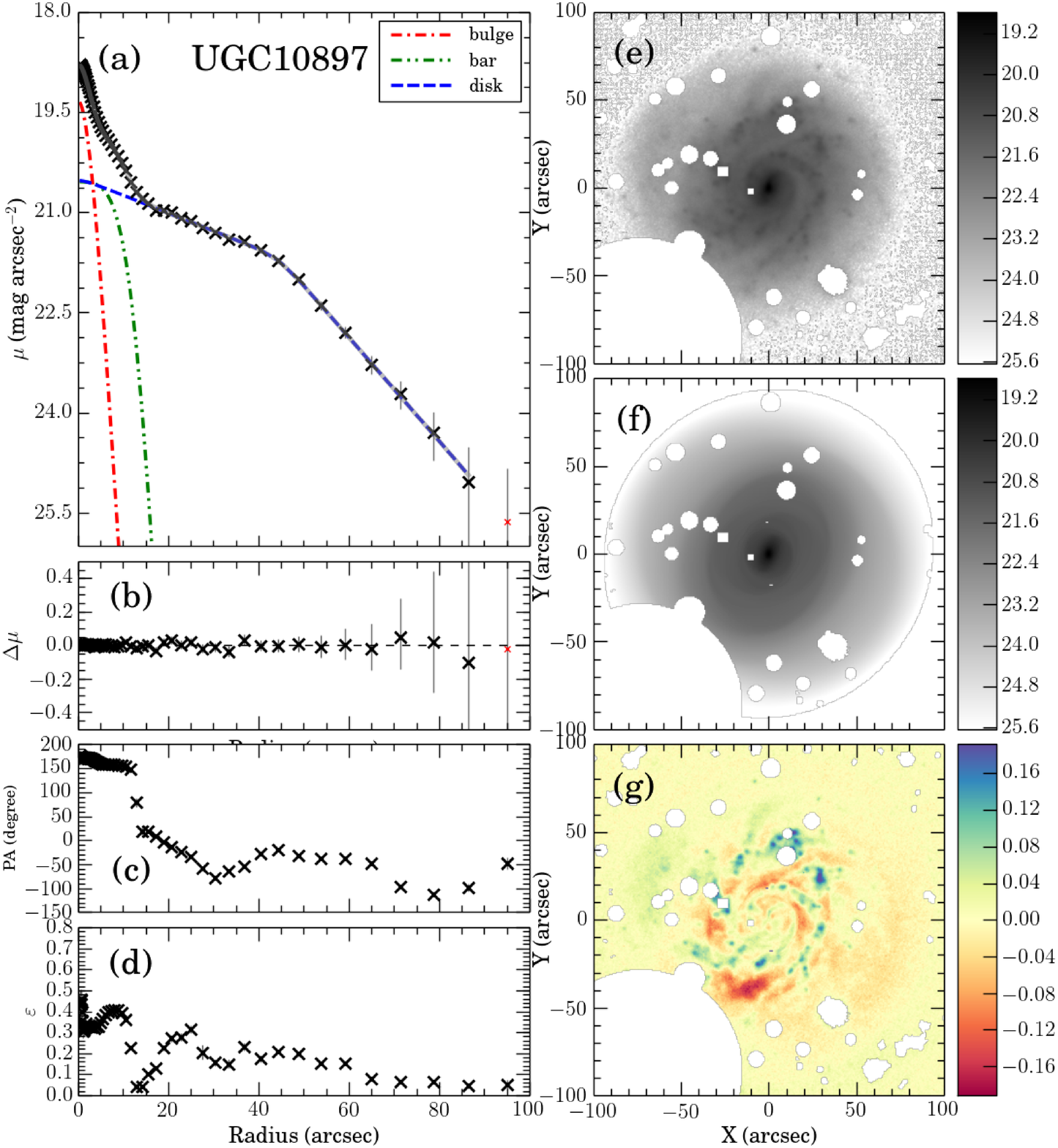}
\caption{Examples of structural decomposition for two galaxies of the GHASP sample in the \Rc-band, UGC 89 and UGC 10897. For each galaxy, we present seven panels, containing (a) surface brightness profiles and their decomposition components, (b) fitting residuals, (c) position angle profile, (d) ellipticity profile, (e) \Rc-band image, (f) \textsc{ellipse} model and (g) residual sigma image. 
} \label{fig:decomp_examples} \end{figure*}

We have performed the decomposition in all \Rc-band profiles in the \ac{ghasp}
OHP sample. We developed a Python routine which performs a weighted chi-square
minimization between the data and a model using the Levenberg-Marquardt
algorithm \citep[see, eg.][]{1992nrfa.book.....P}, using a PSF convolution of
the models with a Moffat function (see section \ref{sec:dataohp}). The input
model is set manually according to the observation of photometric features in
the SB profile and the images of the galaxies. Also, the observation of the
varying ellipticity and position angle as a function of the radius usually
hinted for the different structural sub components of a galaxy. Figure
\ref{fig:decomp_examples} presents examples of structural decomposition of a
sample of twelve galaxies, illustrating the variety of morphologies we have in
our sample and the varied decomposition components we included. The structural
parameters for all galaxies are listed in Table \ref{tab:decomp}.  In the next
section, we give some details on the used parametrizations for the different
components.

\begin{table*} \scriptsize \caption{Decomposition parameters for the \ac{ghasp}
sample in the \Rc band obtained at the OHP observatory. The printed version
contains only a sample of the table, and the remaining material is available
online. (1) Galaxy name. (2-4) Parameters for the S\'ersic function of bulges,
bars and other components. (5) Visual classification of the components. (6-10)
Parameters for the disks according to the broken exponential function. (11)
Classification of the disks regarding to the type of breaks, according to the
scheme of  \citet[][see text for details]{2008AJ....135...20E}. (12) Magnitude
of the central point source.} \begin{tabular}{ccccccccccccc} \hline \hline &
\multicolumn{4}{c}{\underline{\hspace{3cm}S\'ersic\hspace{1.9cm}}}  &
\multicolumn{6}{c}{\underline{\hspace{3.6cm}Disk\vphantom{g}\hspace{4cm}}} &
\underline{\hspace{0.4cm}PS\hspace{0.4cm}}\\ Galaxy     & $\mu_b$ & $r_e$ & $n$
& type & $\mu_d$ & $h_i$ & $h_o$ & $r_b$ & $\alpha$ & Type & $m_{\rm ps}$\\ &
(mag arcsec$^{-2}$) & (arcsec) &  & & & (mag arcsec$^{-2}$) & (arcsec) &
(arcsec) & (arcsec) & & (mag)\\ (1) & (2) & (3) & (4) & (5) & (6) & (7) & (8) &
(9) & (10) & (11) & (12)\\ \hline 
UGC 89 & $16.7\pm0.1$ & $2.3\pm0.1$ & $1.0\pm0.1$ & bulge & $21.5\pm1.0$ & $35\pm25$ & --- & --- & --- & Type I & ---\\
 & $20.3\pm0.2$ & $19\pm1$ & $0.2\pm0.1$ & bar & --- & --- & --- & --- & --- &  & ---\\
 & $23\pm1$ & $29\pm5$ & $0.11\pm0.07$ & bar & --- & --- & --- & --- & --- &  & ---\\
UGC 94 & $18.7\pm0.3$ & $0.8\pm0.3$ & $1.6\pm0.3$ & bulge & $20.3\pm0.1$ & $14.0\pm0.8$ & --- & --- & --- & Type I & ---\\
 & $20.79\pm0.09$ & $8.3\pm0.4$ & $0.14\pm0.09$ & bar & --- & --- & --- & --- & --- &  & ---\\
UGC 508 & $18.6\pm0.2$ & $3.4\pm0.1$ & $2.65\pm0.04$ & bulge & $19.60\pm0.04$ & $26.2\pm0.2$ & --- & --- & --- & Type I & ---\\
 & $23.69\pm0.08$ & $20.2\pm0.3$ & $0.05\pm0.05$ & bar & --- & --- & --- & --- & --- &  & ---\\
 & $21.19\pm0.06$ & $7.6\pm0.3$ & $0.22\pm0.02$ & bar & --- & --- & --- & --- & --- &  & ---\\
UGC 528 & $19.0\pm0.2$ & $5.4\pm0.6$ & $2.7\pm0.3$ & bulge & $18.6\pm0.2$ & $19\pm2$ & --- & --- & --- & Type I & ---\\
 & $19.8\pm0.2$ & $13.1\pm0.8$ & $0.1\pm0.2$ & arms & --- & --- & --- & --- & --- &  & ---\\
 & $20.3\pm0.4$ & $22\pm1$ & $0.15\pm0.09$ & arms & --- & --- & --- & --- & --- &  & ---\\
UGC 763 & $28.0\pm0.3$ & $311\pm183$ & $5.3\pm0.3$ & bulge & $19.99\pm0.06$ & $25\pm2$ & --- & --- & --- & Type I & ---\\
 & $23\pm2$ & $18\pm9$ & $0.05\pm0.06$ & bar & --- & --- & --- & --- & --- &  & ---\\
 
... & ... & ... & ... & ... & ... & ... & ... & ... & ... & ... & ...\\
 
 UGC 12276 & $21.1\pm0.1$ & $5.4\pm0.2$ & $2.56\pm0.01$ & bulge & $20.24\pm0.09$ & $14.0\pm0.1$ & $7.0\pm0.3$ & $45.0\pm0.1$ & $1.0\pm0.2$ & Type II & ---\\
 & $27.36\pm0.05$ & $0.1\pm0.1$ & $6.59\pm0.04$ & bar & --- & --- & --- & --- & --- &  & ---\\
UGC 12276c & $22.74\pm0.09$ & $0.0\pm0.1$ & $2.99\pm0.08$ & bulge & $20.60\pm0.09$ & $3.6\pm0.2$ & --- & --- & --- & Type I & ---\\
UGC 12343 & $20.9\pm0.4$ & $12\pm3$ & $2.9\pm0.3$ & bulge & $19.2\pm0.1$ & $30\pm3$ & --- & --- & --- & Type I & ---\\
 & $21.7\pm0.4$ & $56\pm3$ & $0.2\pm0.1$ & bar & --- & --- & --- & --- & --- &  & ---\\
UGC 12632 & $23.1\pm0.2$ & $2.9\pm0.6$ & $0.05\pm0.08$ & nucleus & $22.1\pm0.1$ & $58\pm4$ & --- & --- & --- & Type I & ---\\
 & $23.4\pm0.2$ & $9.8\pm0.4$ & $0.10\pm0.02$ & bulge & --- & --- & --- & --- & --- &  & ---\\
UGC 12754 & $20.36\pm0.07$ & $11.8\pm0.4$ & $0.43\pm0.04$ & bar & $20.6\pm0.2$ & $56\pm7$ & --- & --- & --- & Type I & $15.7\pm0.2$\\
 & $22.4\pm0.4$ & $24\pm1$ & $0.05\pm0.06$ & bar & --- & --- & --- & --- & --- &  & ---\\
 \hline \hline \end{tabular} \label{tab:decomp} \normalsize
\end{table*}

\subsubsection{Disks}

Since early works of \citet{1940BHarO.914....9P} and
\citet{1958ApJ...128..465D}, the intensity profile of disks have been mostly
described by a simple exponential law, 

\begin{equation} I_d(r)=I_0\exp\left(-\frac{r}{h}\right), \label{eq:exponential}
\end{equation}

\noindent where $I_0$ is the central ($r=0$) intensity of the disk and $h$ is
the disk scale length. Usually, we refer to the central intensity in terms of
surface brightness using the relation $\mu_0=-2.5\log I_0$. For the case of
exponential disks, the  total apparent magnitude is given by

\begin{equation} m_{\text{disk}} = -2.5 \log \left (2\pi I_0
h^2\frac{b}{a}\right ) \mbox{,} \label{eq:mdisk} \end{equation}

\noindent where $b/a=1-\varepsilon$ is the minor-to-major axis ratio of the
galaxy.

However, deviations from a simple exponential disk were already noticed by
\citet{1970ApJ...160..811F}, specially in the form of truncations or breaks in
the inner profiles of galaxies. Later,  \citet{1982A&A...110...61V} also noticed
breaks at large radii of disks, and more recently deviations at very low surface
brightness have been observed, including upward bends
\citep[e.g.][]{2005ApJ...626L..81E}.

Based on these observations, \citet{2008AJ....135...20E} proposed a reviewed
classification of disks in three categories: Type I, simple disks well described
by exponential disks; Type II, disks with downward truncations; and Type III, 
disks with upward bends. A local census of disk properties performed by
\citet{2006A&A...454..759P} of late-type galaxies (Sb-Sm) estimated the fraction
of Type I galaxies to be only 10\%, while 60\% are classified as Type II and
30\% are Type III according to this new classification scheme. Therefore, an
updated profile for the disks is here adopted whenever breaks are clearly
observed, by using broken exponential profiles, given by
\citep{2008AJ....135...20E}

\begin{equation} I_d(r)=SI_0\exp \left(-\frac{r}{h_i}\right) \cdot\left\{1+\exp
\left[\alpha (r - r_b)\right ]
\right\}^{\frac{1}{\alpha}(\frac{1}{h_i}-\frac{1}{h_o})}\mbox{,}
\label{eq:brokenexp} \end{equation}

\noindent where $I_0$ is the central intensity of the disk, $h_i$ and $h_o$ are
the inner and outer disk scale length respectively, $r_b$ is the break radius,
$\alpha$ is the sharpness of the disk transition between the inner and outer region
(where low $\alpha$ means a smooth transition
from the inner to the outer disk and high $\alpha$ means an abrupt transition), and
$S$ is a scaling factor given by 

\begin{equation} S=\left[1+\exp (-\alpha
r_b)\right]^{-\frac{1}{\alpha}(\frac{1}{h_i}-\frac{1}{h_o})}\mbox{.}
\end{equation}

\noindent For the broken disks in our sample, the total luminosities were 
calculated numerically, given that a solution by the integration
of equation \eqref{eq:brokenexp} is beyond the scope of this work. 

In Table \ref{tab:decomp}, we include a classification of the disks according to
our observations. However, it is important to notice that breaks may occur at
different radial distances, with different physical interpretations: inner
breaks ($\mu_r\sim23$ mag arcsec$^{-2}$) may be related to star formation, while
outer breaks ($\mu_r\sim27$ mag arcsec$^{-2}$) may indicate a real drop in the
stellar mass density \citep{2012MNRAS.427.1102M}. Therefore, our classifications
are restricted to the mean limiting surface brightness of 24.5 mag
arcsec$^{-2}$. 

\subsubsection{Other components}

Apart from the disk, several other components are observed, including bulges,
bars, arms, rings and lenses. We included those components in the decomposition
using a S\'ersic function \citep{sersic68}, given by 

\begin{equation} I_b(r)=I_e\exp \left ( -b_n \left [ \left ( \frac{r}{r_e}\right
)^{1/n}-1\right ]\right ), \label{eq:sersic} \end{equation}

\noindent where $r_e$ is the effective scale of the component (for which 50\% of
the light is within $r_e$), $I_e$ is the intensity at the effective radius, and
$n$ is the S\'ersic index. The term $b_n$ is not a free parameter, but a
function of the S\' ersic index due to the parametrization of the function at
the effective radius instead of at the centre. In our calculations, we adopted
the expressions for $b_n$ presented in the  Appendix [A1] of
\citet{2003ApJ...582..689M}. In a first order approximation, $b_n\approx
2n-0.33$, although the error may be considerable for $n<0.5$. For
simplification, we also rescale the effective intensity to surface brightness
using the expression $\mu_e=-2.5\log I_e$. In the case of equation
\eqref{eq:sersic}, the total magnitude is given by
\citep{1991A&A...249...99C,2003ApJ...582..689M}

\begin{equation} m_{\rm sersic}=-2.5\log \left (\frac{2\pi I_e r_e^2
\mathrm{e}^{b_n} n \Gamma(2n)}{ b_n^{2n}} \frac{b}{a}\right)\mbox{,}
\label{eq:msersic} \end{equation} 

\noindent where $\Gamma(x)$ is the complete gamma function of a variable $x$.
The S\'ersic profile is a generalization of other commonly used profile
functions, such as the exponential for $n=1$, a Gaussian for $n=\frac{1}{2}$ and
a \citeauthor{1948AnAp...11..247D}'s profile \citep{1948AnAp...11..247D} for
$n=4$.  Moreover, the S\'ersic index can also be used as an indicator of the
kind of component that is being observed. For example, in the optical and
near-infrared wavelengths, \citet{2010ApJ...716..942F} have shown that $n\lesssim2$
may indicate a pseudobulge, whereas $n\gtrsim 2$ may indicate a classic bulge
for the spheroidal components. Besides, bars have typically $n\sim0.7$
\citep{2011MNRAS.415.3308G}. In the Table \ref{tab:decomp}, where the
decomposition results are presented, we include a classification of the S\'ersic
function components according to the visual inspection of images and profiles,
such as bulges, bars, lenses and spiral arms. 

In 35 galaxies, a nuclear source is also detected, which may be related to
different physical processes, such as an active nucleus or a stellar
concentration. We have tested two approaches for parametrizing these components, using 
either a S\'ersic component or a single delta function with a peak at 
$r=0$. In 12 cases, the former approach resulted in a better description 
of the nucleus, because they have slightly larger FWHM than that of the 
modelled PSF and/or because of the different shape of the nuclear 
source compared to a star. These components are described as nucleus in tables 
\ref{tab:decomp} in the column 5. For 23 galaxies, however, the later approach of using a delta function
resulted in a better description of the nucleus in these cases. This delta function has only one free parameter, the magnitude of 
the source ($m_{\rm ps}$), and its profile is of a field star which is 
described as a Moffat function.  These point source magnitudes are included 
in the last row of Table \ref{tab:decomp}.  

\section{Photometric Internal Consistency and Literature Comparison}
\label{sec:tests}

\begin{figure*} \centering
\includegraphics[width=\textwidth]{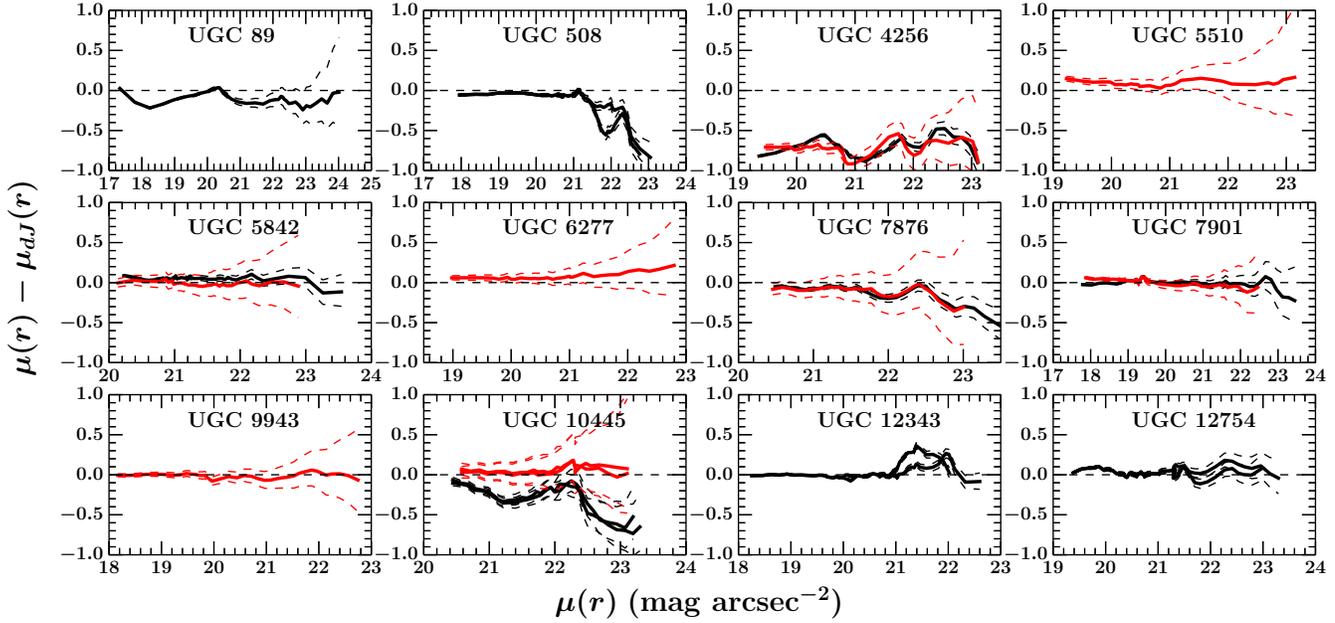} \caption{Difference
between the surface brightness as function of the isophotal level for 12
galaxies profiles in common with de Jong and van der Kruit (1994) in the
\Rc-band. The solid black (red) lines represent the data from the \ac{ohp}
(\ac{sdss}), while the dashed lines are the profile errors. Multiple profiles of
galaxies in \citet{1994A&AS..106..451D} are shown in different lines. Dashed
lines represent the errors in our profiles.} \label{fig:compdejong}
\end{figure*}

In this section we make a series of tests on our photometric results to verify
their consistency and to compare them with similar results in the literature. We
have already made an internal consistency check of our SB profiles in Figure
\ref{fig:ugriztoR}, where we observed that the SB profiles from \ac{sdss} data
are similar to those observed with direct measurements in the \Rc-band. Our SB
profiles can also be compared with those derived for twelve galaxies in common with
\citet{1994A&AS..106..451D} in the \Rc-band, as shown in Figure
\ref{fig:compdejong}, where black and red lines show the difference between our
and the literature data for the \ac{ohp} and \ac{sdss} data sets respectively.
To make a proper comparison, we have fixed the position angle and the
ellipticity of the galaxies to mimic the method of those authors instead of
using free position angle and ellipticity as in section \ref{sec:profiles}. We also limited the
comparison to regions greater than the seeing of our images. 

The most deviant case is UGC 4256, but the internal consistency of our results
for two different data sets indicates a possible systematic offset in the data
of \citet{1994A&AS..106..451D} for this galaxy. The deviation in the outer
region of UGC 508 can be explained by the limited field of view in the images of
\citet{1994A&AS..106..451D} which cuts part of the galaxy. In this case, it is
also possible to see a large variation of the isophotes fainter than 21 mag
arcsec$^{\rm2}$ within the observations of \citeauthor{1994A&AS..106..451D}.
Finally, the problematic case of UGC 10445 for the \ac{ohp} data can be
explained by the relatively short exposure time for this object, including only
one image, which affects the accuracy of the sky subtraction. Apart from these
remarks, the overall picture is a good agreement with
\citet{1994A&AS..106..451D}, which is in turn in agreement with several other
authors in the literature.  

\begin{figure*} \centering
\includegraphics[width=0.98\textwidth]{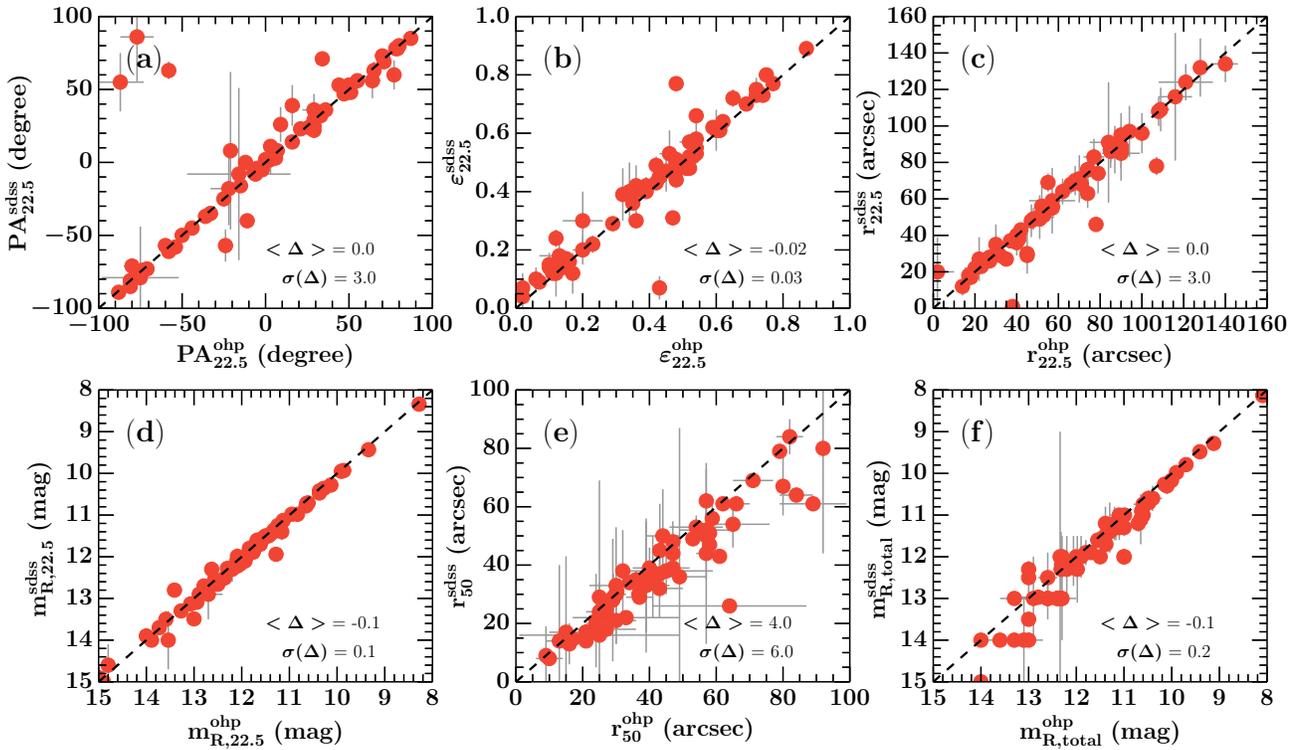}
\caption{Comparison of photometric properties derived for the 66 galaxies in
common between the \ac{ohp} and \ac{sdss} datasets. The parameter $\Delta$
represents the difference between the parameters in the two data sets, with its
the mean value ($<\Delta>$) and its standard deviation ($\sigma (\Delta)$) shown
in the bottom of each panel. Parameters are displayed as the following: (a)
isophotal position angle; (b) isophotal ellipticity; (c) isophotal radius; (d)
isophotal apparent magnitude; (e) effective radius; (f) total apparent
magnitude. The horizontal and vertical axes display the results for the \ac{ohp}
and \ac{sdss} datasets respectively, and the dashed line represents the
ideal equality between datasets.} \label{fig:internalcheck} \end{figure*}

In Figure \ref{fig:internalcheck}, we show the internal consistency of the
isophotal and integrated photometric parameters derived from the SB profiles by
the comparison of the results obtained with the \ac{ohp} and \ac{sdss} data
sets, including position angle, ellipticity, integrated magnitude and radius at
the isophote of 22.5 mag/arcsec$^2$ and also effective radius and total
magnitude. We also display the mean residual difference ($<\Delta>$) and its
standard deviation ($\sigma (\Delta)$) for each parameter, resulting in
compatible measurements in both data sets within one standard deviation.

\begin{figure} \centering
\includegraphics[width=0.48\textwidth]{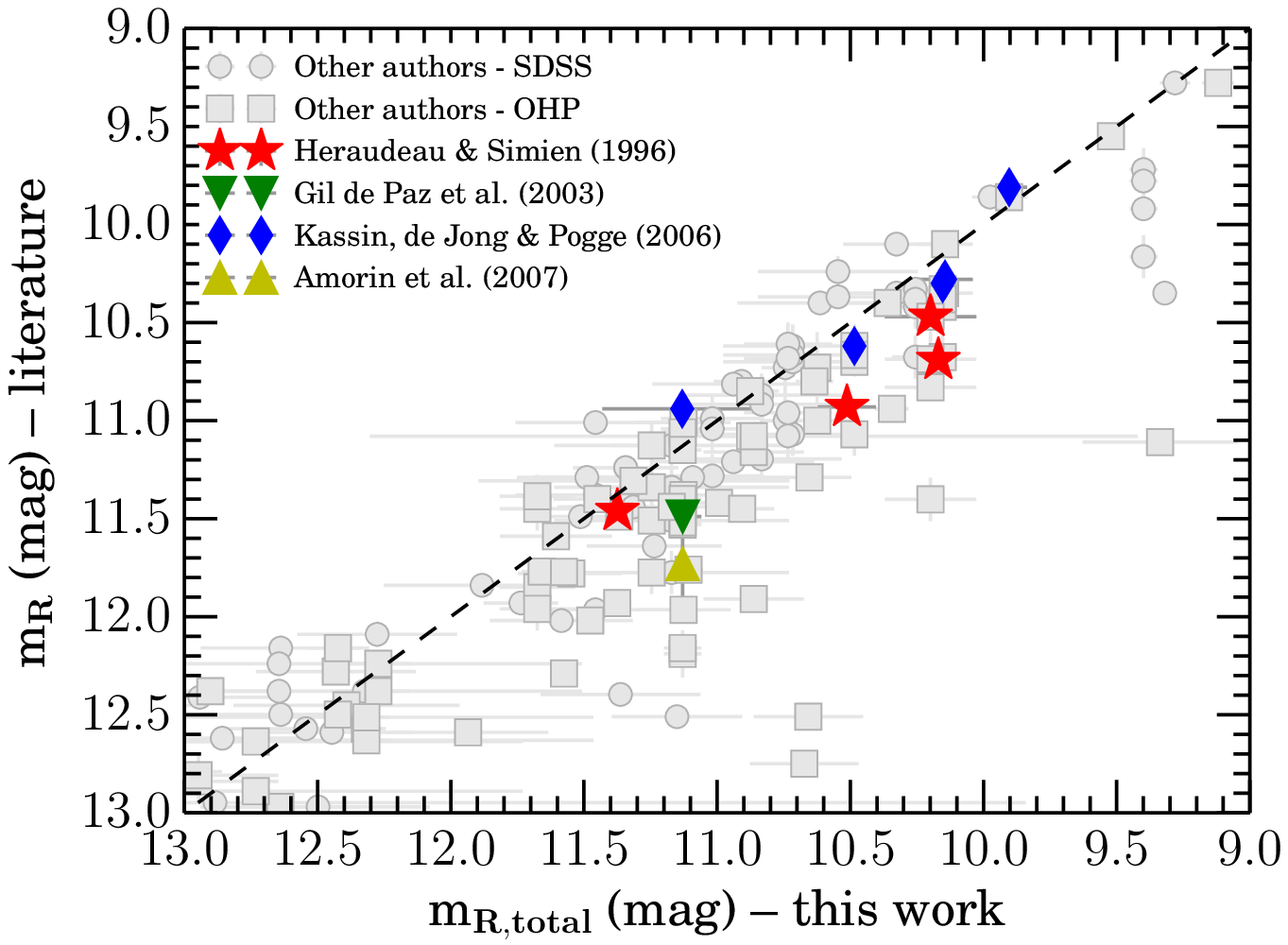}
\caption{Comparison of total apparent magnitudes with the literature.
Coloured symbols indicate the comparisons with data in the \Rc-band, including
the works of  \citet{1996A&AS..118..111H}, \citet{2003ApJS..147...29G},
\citet{2006ApJS..162...80K} and \citet{2007A&A...467..541A}. Gray symbols
indicate the comparisons with magnitudes in pass bands similar to \Rc with our
magnitudes both for the \ac{sdss} (circles) and \ac{ohp} (crosses) samples. This
later data include magnitudes from \citet{1989ApJS...70..687H},
\citet{1996AJ....112.2471T},  \citet{2004A&A...414...23J},
\citet{2004AJ....127.1386C},  \citet{2005ApJ...630..784T},
\citet{2005MNRAS.361...34D},  \citet{2007AJ....134.2286H},
\citet{2007MNRAS.376.1480N},  \citet{2008A&A...486..755T},
\citet{2008A&A...487..485H},  \citet{2008AJ....135..291M} and
\citet{2011A&A...527A.101K}. The dashed line represents equality between
measurements.} \label{fig:magcomp} \end{figure}

In Figure \ref{fig:magcomp} we compare our total magnitudes with data in the
literature. There are just a few works in the literature for which total
magnitudes are measured in the \Rc-band, so we also include in the figure
measurements in similar filters in the literature without any additional
correction or extrapolation, which are displayed as gray symbols, including
\citet{1996AJ....112.2471T},  \citet{2004A&A...414...23J},
\citet{2004AJ....127.1386C},  \citet{2005ApJ...630..784T},
\citet{2005MNRAS.361...34D},  \citet{2007AJ....134.2286H},
\citet{2008A&A...486..755T},  \citet{2008A&A...487..485H},
\citet{2008AJ....135..291M} and   \citet{2011A&A...527A.101K}. However,
specially relevant is the comparison with proper \Rc magnitudes, which we
highlight in Figure \ref{fig:magcomp} using coloured symbols for the works of
\citet{1996A&AS..118..111H}, \citet{2003ApJS..147...29G},
\citet{2006ApJS..162...80K} and \citet{2007A&A...467..541A}. The number of
overlapping galaxies with \Rc-band data is scarce, only 12 galaxies, but those
are in good agreement with most previous works, specially with the more recent
survey of \citet{2006ApJS..162...80K}. 

\begin{figure} \centering
\includegraphics[width=0.44\textwidth]{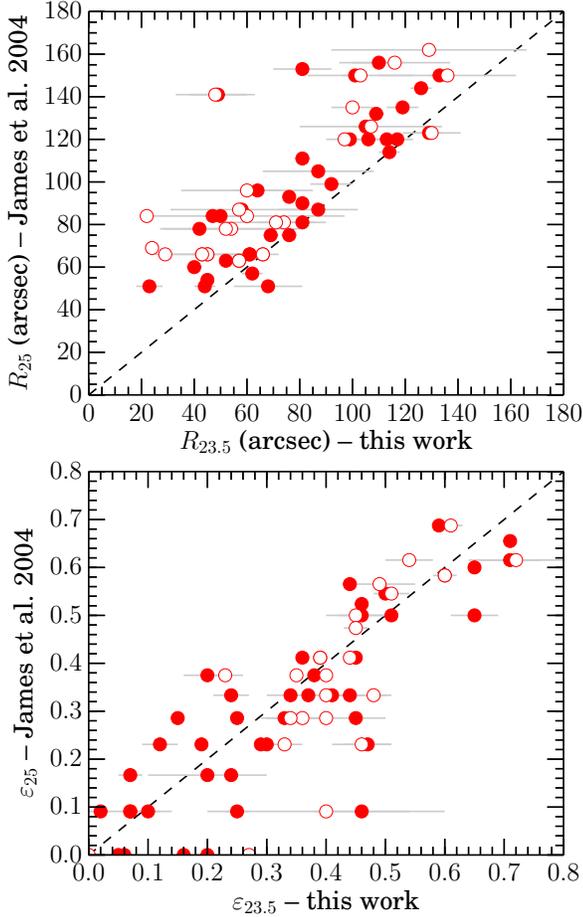} \caption{Comparison of
our isophotal radius (above) and ellipticity (below) with those derived by 
\citet{2004A&A...414...23J}. Filled (hollow) points represent data from OHP
(SDSS), and the dashed lines represent the equality between measurements.}
\label{fig:diameters} \end{figure}

\begin{figure} \centering
\includegraphics[width=0.44\textwidth]{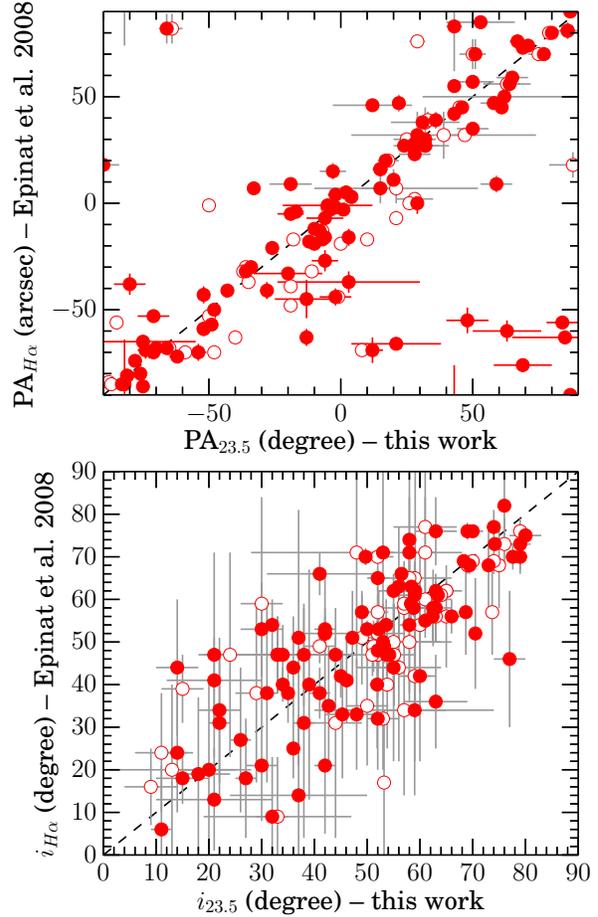} \caption{Comparison
of our isophotal position angles (above) and inclinations (below) with those derived from the
H$\alpha$ velocity fields of Epinat et al. 2008. Filled (hollow) points
represent data from OHP (SDSS), and the dashed line represents the equality
between measurements.} \label{fig:inclinations} \end{figure}

In the top panel of Figure \ref{fig:diameters}, we compare our isophotal radius
with the results of \citet{2004A&A...414...23J}. As the reference  isophotal
levels are different, there is a systematic offset in the isophotal radius in
the sense that results from our work are systematically smaller than those from
\citeauthor{2004A&A...414...23J}, but still there is a good correspondence
between the two datasets. In the bottom panel of Figure \ref{fig:diameters}, we
show that the ellipticities are also well correlated, as expected, because at
both reference isophotal levels the disk is the dominant component in the light
of the galaxy, and has a simple geometry that does not vary drastically between
these pass bands. Finally, in figure \ref{fig:inclinations}, we compare the
isophotal position angles and inclinations with the results from the
H$\alpha$ map analysis of Epinat et al. 2008, which demonstrate that our
analysis produces results that are similar even to other tracers of the galaxy
shape such as the gas. 

\section{Scaling relations} \label{sec:scaling}

Scaling relations contain important information about the physical processes
regarding galaxy formation and evolution, and impose important constraints to
models that attempt to describe such objects \citep{2007ApJ...671..203C}. In
this section, we derive the most significant scaling relation involving 
luminosities, sizes and rotation curve velocity for each of the two most
basic structural components of the galaxy, the bulge and the disk, and also for
the whole galaxy. In section \ref{subsec:corrections}, we show how we correct 
the sizes and luminosities for the effects of distance, inclination and 
dust attenuation, and in section \ref{subsec:fitting} we show how we 
estimate the scaling relations. In section  \ref{subsec:corr} we present
the main results and in section \ref{sec:tf} we explore the Tully-Fisher 
relation in greater detail.

\subsection{Correction for the effects of inclination and distances}
\label{subsec:corrections}

We use all galaxies in table \ref{tab:decomp} with a bulge component according
to our classification in the decomposition. Disk apparent magnitudes
($m_{\text{disk}}$) are calculated using equation \eqref{eq:mdisk} or by
numerical integration in the case of broken profiles for the disks, while bulge
luminosities ($m_{\text{bulge}}$) are calculated using equation
\eqref{eq:sersic}. The absolute magnitudes of these components are then obtained
using the equations

\begin{eqnarray} M_{\text{disk}}  = m_{\text{disk}} - d_1 - d_2 (1 - \cos
i)^{d_3}-5\log D -25\mbox{,}\\ M_{\text{bulge}} = m_{\text{bulge}} - b_1 - b_2
(1 - \cos i)^{b_3}-5\log D -25\mbox{,} \end{eqnarray}

\noindent where the internal extinction coefficients $b_1=0.60$, $b_2=1.33$,
$b_3=1.75$, $d_1=0.15$, $d_2=1.09$ and $d_3=2.82$ are obtained by linear
interpolation from Table 1 of \citet{2008ApJ...678L.101D} for the \Rc-band
($\lambda=647$nm), $D$ is the distance in Mpc according to \citet{2008MNRAS.388..500E},
and the inclination $i$ is taken from the gas velocity field analysis in
\citet{2008MNRAS.388..500E}, if available, or from the isophotal analysis
otherwise. Individual distance errors are rarely available, and we adopt a
value of $25\%$ for all objects.

The total luminosity of each galaxy is estimated by its total magnitude
according to the analysis of the curve of growth of the SB profiles (see section
\ref{sec:intphot}). In this case, we obtain the absolute magnitude of the
galaxies ($M_{\rm R,total}$) from the apparent total magnitudes ($m_{\rm
R,total}$) using the expression  

\begin{equation} M_{\rm R,total}=m_{\rm R, total}-A_{i}(R)-5\log D -25\text{,}
\end{equation}

\noindent where $A_{i}(R)$ is the internal extinction correction given by
\citet{1998AJ....115.2264T}

\begin{equation} A_{i}(R)=\log (b/a) \times \left \{ 1.15 + 1.88\left ( \log
2V_{\text{max}} - 2.5\right ) \right \}\text{.} \end{equation}

\noindent Here $b/a$ is again the minor-to-major axis ratio and $V_{\text{max}}$
is the maximum velocity of the H$\alpha$ rotation curve derived from the
velocity field analysis from  \citet{2008MNRAS.388..500E}. 

We compare these luminosities with the physical sizes of each component, using
the scale length of disks ($h$), or the inner disk length in the cases of broken
disks, the effective radius of the bulge ($r_e$) and the effective radius of the
galaxy ($r_{50}$). We do not attempt to correct the sizes of the components for 
extinction, and we only rescale the sizes according to the distance. 
Finally, we use the velocity $V_{\text{max}}$ as in
\citet{2008MNRAS.388..500E} as our dynamical tracer, excluding galaxies for
which the flat part of the rotation curve was not reached in the velocity field
analysis of the H$\alpha$ observations.

\subsection{Fitting method} \label{subsec:fitting}

We assessed the statistical significance of pairs of luminosities, sizes and
velocities using the Spearman's rank correlation coefficient $r$, which is a
measurement of the strength of the correlation of two variables, and the
associated p-value $p$, which indicates the probability of obtaining a result at
least as extreme as the one obtained from a random distribution, both indicated
in the boxes at the top of each panel. For 18 cases, we obtained correlations
with $p<0.1$\%, for which we calculated scaling laws considering the direct and inverse
cases. The relations, displayed in the form of dashed lines in Figure \ref{fig:corr}, 
were calculated as in the following. We consider a linear relation in the form of 

\begin{equation} y_i = \alpha x_i + \beta\mbox{,} \label{eq:linear}
\end{equation}

\noindent where each galaxy is represented by the index $i$, $\alpha$ is the
slope of the relation and $\beta$ is the zero point. We then performed a
$\chi^2$ minimization considering the measurement uncertainties in both
variables, considering also an intrinsic scatter, $\varepsilon_0$, for the
relation \citep[see][]{2002ApJ...574..740T}, using the relation

\begin{equation} \chi^2_\nu=\frac{1}{N-2}\sum_{i=1}^{N}\frac{\left(y_i - \beta -
\alpha x_i\right
)^2}{\varepsilon_{yi}^2+\alpha^2\varepsilon_{xi}^2+\varepsilon_0^2}\mbox{,}
\label{eq:chi2} \end{equation}

\noindent where $N$ is the number of galaxies of the sample, $\nu=N-2$ is the
number of degrees of freedom, $\varepsilon_{xi}$ and $\varepsilon_{yi}$ are the
parameter uncertainties. The presence of the variables in both the numerator and
in the denominator of relation \eqref{eq:chi2} makes the equation non-linear,
and most common methods of minimization, such as the Levenberg-Marquardt
algorithm \citep[][]{1992nrfa.book.....P}, may have problems to obtain
convergence. To obtain stable solutions, we used the interactive method
described by \citet{2006MNRAS.373.1125B}, which consists in solving equation
\eqref{eq:chi2} for a fixed value of $\varepsilon_0$, and then update
$\varepsilon_0$ by multiplying for $\chi^2_\nu$ elevated to a power of $2/3$,
until obtaining $\chi^2_\nu=1$. This method failed in only two cases, as 
show in table \ref{tab:corr}, because either the dispersion is too low and 
$\chi^2<1$ or if the dispersion was too high. The uncertainties for the 
coefficients were estimated using the bootstrapping method.  

\subsection{Results} \label{subsec:corr}

\begin{figure*} \centering \includegraphics[width=\textwidth]{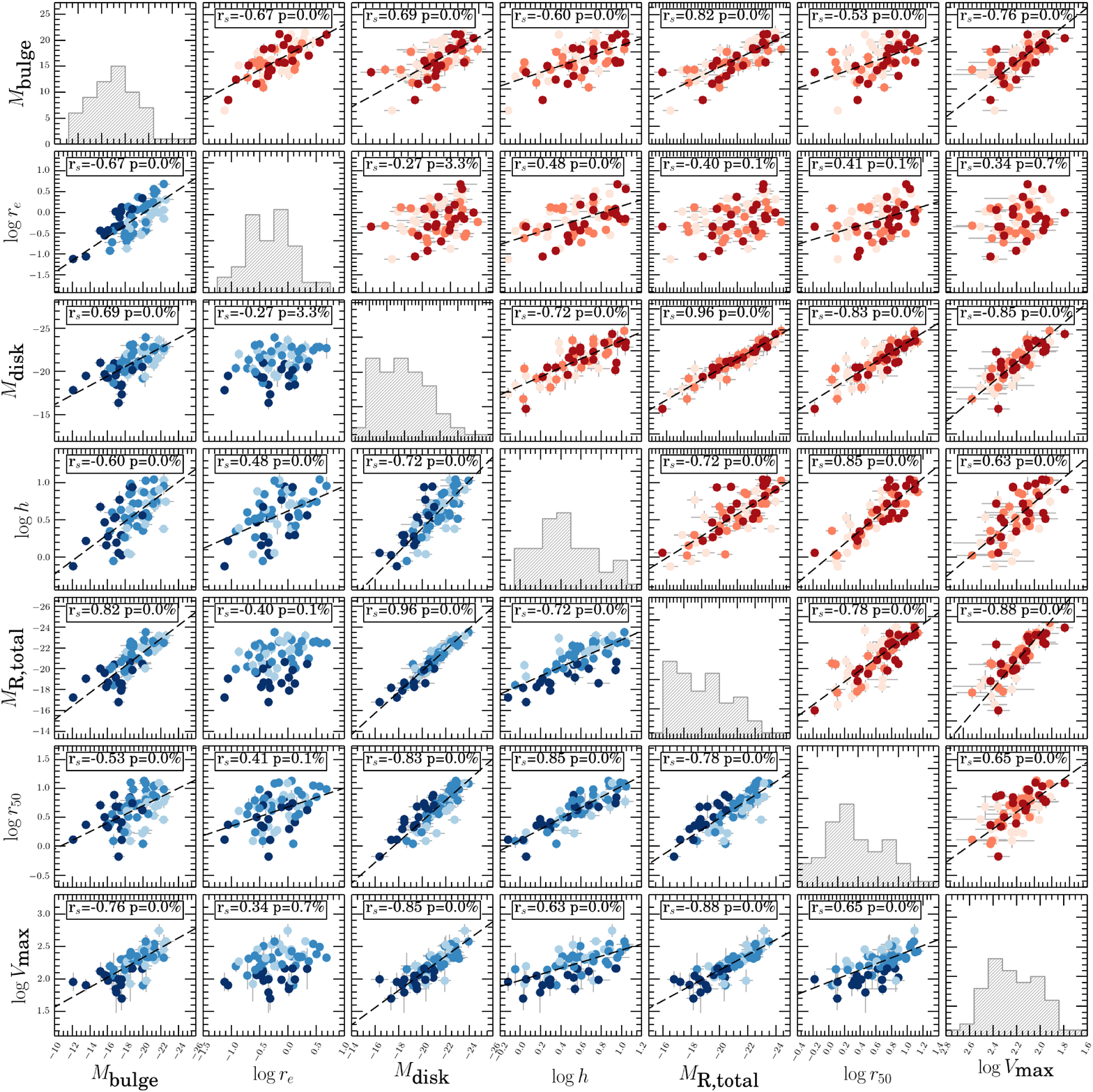}
\includegraphics[width=\textwidth]{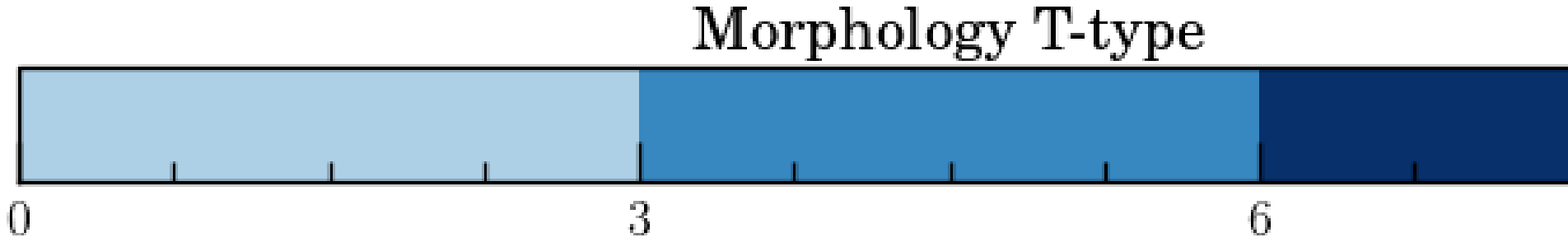} \caption{\Rc-band
correlations between luminosities and sizes of bulges ($M_{\rm bulge}$, $\log
r_e$), disks ($M_{\rm disk}$ and $\log h$) and the galaxy luminosities ($M_{\rm
R,total}$), sizes ($r_{50}$) and  maximum velocity of the rotation curve
($V_{\rm max}$). Panels above the diagonal display colours of galaxies according
to the colour map in the bottom right of the figure, which divide galaxies
according to the presence of bars following the morphological classification.
Panels below the diagonal display colours according to the colourmap in the bottom
left of the figure, classifying galaxies according to their numerical type in
the de Vaucouleurs's classification. The Spearman's rank correlation coefficient
($r_s$) and the p-value of the relation are presented in the box on top of each
panel. For relations with significance greater than 3$\sigma$, we include a
dashed black line to indicate the best fit linear regression, whose coefficients
are presented in Table \ref{tab:corr}. The diagonal panels display the
distributions of parameters as histograms.} \label{fig:corr} \end{figure*}

Figure \ref{fig:corr} shows the resulting relations among luminosities, sizes
and velocities for 62 galaxies selected in the previous section. We 
adopt different colouring for the panels above and below
the diagonal according to the strength of the bar and to the morphological
classification respectively. However, we could not study these correlations in
subsamples due to the low statistics after dividing the data in those classes,
and our quantitative results are all related to the complete photometric sample.
The summary of the scaling laws is shown in table \ref{tab:corr}, where we sort
the relations by decreasing Spearman's coefficients.

\begin{table*} \caption{\label{tab:corr}Scaling relations with statistical
significance above 3$\sigma$ for the luminosities, sizes and velocity of the
galaxies and its two basic subcomponents, bulges and disks, ranked by decreasing
Spearman's rank coefficients ($r_s$). The first column indicates the identification of
the relation. The second column indicates the variables involved in the relation,
as well as the correlation coefficients and their p-values. The third column
indicates the direct and inverse relations. The fourth column shows the intrinsic
scatter of the relation.} \begin{tabular}{cclcc} \hline ID & Parameters &
Relation & $\varepsilon_0$\\ (1) & (2) & (3) & (4)\\ \hline 
 (a) & $M_{\mbox{R,total}}$-$M_{\mbox{disk}}$ &  $M_{\mbox{disk}}=(0.98\pm0.03)M_{\mbox{R,total}}+(-0.6\pm0.7)$ & --- \\
 & $|r|=0.96$, $p=3\cdot 10^{-34}$ &  $M_{\mbox{R,total}}=(1.02\pm0.04)M_{\mbox{disk}}+(0.6\pm0.8)$ & --- \\ \\
 (b) & $\log V_{\mbox{max}}$-$M_{\mbox{R,total}}$ &  $M_{\mbox{R,total}}=(-8.0\pm0.7)\log V_{\mbox{max}}+(-3\pm1)$ & $0.37\pm0.05$ \\
 & $|r|=0.88$, $p=6\cdot 10^{-21}$ &  $\log V_{\mbox{max}}=(-0.12\pm0.01)M_{\mbox{R,total}}+(-0.2\pm0.2)$ & $0.04\pm0.01$ \\ \\
 (c) & $\log V_{\mbox{max}}$-$M_{\mbox{disk}}$ &  $M_{\mbox{disk}}=(-7.1\pm0.8)\log V_{\mbox{max}}+(-5\pm2)$ & $0.31\pm0.07$ \\
 & $|r|=0.85$, $p=2\cdot 10^{-18}$ &  $\log V_{\mbox{max}}=(-0.13\pm0.01)M_{\mbox{disk}}+(-0.6\pm0.3)$ & $0.04\pm0.01$ \\ \\
 (d) & $\log r_{50}$-$\log h$ &  $\log h=(0.88\pm0.07)\log r_{50}+(0.00\pm0.04)$ & $0.132\pm0.003$ \\
 & $|r|=0.85$, $p=4\cdot 10^{-18}$ &  $\log r_{50}=(0.93\pm0.06)\log h+(0.11\pm0.03)$ & $0.135\pm0.004$ \\ \\
 (e) & $\log r_{50}$-$M_{\mbox{disk}}$ &  $M_{\mbox{disk}}=(-4.7\pm0.3)\log r_{50}+(-18.1\pm0.2)$ & $0.61\pm0.03$ \\
 & $|r|=0.83$, $p=9\cdot 10^{-17}$ &  $\log r_{50}=(-0.18\pm0.02)M_{\mbox{disk}}+(-3.2\pm0.3)$ & $0.12\pm0.01$ \\ \\
 (f) & $M_{\mbox{R,total}}$-$M_{\mbox{bulge}}$ &  $M_{\mbox{bulge}}=(1.1\pm0.1)M_{\mbox{R,total}}+(5\pm3)$ & $1.11\pm0.02$ \\
 & $|r|=0.82$, $p=2\cdot 10^{-16}$ &  $M_{\mbox{R,total}}=(0.65\pm0.06)M_{\mbox{bulge}}+(-9\pm1)$ & $0.84\pm0.02$ \\ \\
 (g) & $\log r_{50}$-$M_{\mbox{R,total}}$ &  $M_{\mbox{R,total}}=(-4.4\pm0.5)\log r_{50}+(-18.1\pm0.3)$ & $0.88\pm0.02$ \\
 & $|r|=0.78$, $p=9\cdot 10^{-14}$ &  $\log r_{50}=(-0.16\pm0.02)M_{\mbox{R,total}}+(-2.7\pm0.3)$ & $0.167\pm0.005$ \\ \\
 (h) & $\log V_{\mbox{max}}$-$M_{\mbox{bulge}}$ &  $M_{\mbox{bulge}}=(-9\pm1)\log V_{\mbox{max}}+(2\pm2)$ & $1.17\pm0.07$ \\
 & $|r|=0.76$, $p=7\cdot 10^{-13}$ &  $\log V_{\mbox{max}}=(-0.08\pm0.01)M_{\mbox{bulge}}+(0.8\pm0.2)$ & $0.107\pm0.007$ \\ \\
 (i) & $\log h$-$M_{\mbox{disk}}$ &  $M_{\mbox{disk}}=(-4.3\pm0.4)\log h+(-18.6\pm0.2)$ & $0.86\pm0.02$ \\
 & $|r|=0.72$, $p=3\cdot 10^{-11}$ &  $\log h=(-0.16\pm0.02)M_{\mbox{disk}}+(-2.8\pm0.4)$ & $0.165\pm0.006$ \\ \\
 (j) & $M_{\mbox{R,total}}$-$\log h$ &  $\log h=(-0.15\pm0.02)M_{\mbox{R,total}}+(-2.5\pm0.3)$ & $0.181\pm0.004$ \\
 & $|r|=0.72$, $p=5\cdot 10^{-11}$ &  $M_{\mbox{R,total}}=(-4.3\pm0.6)\log h+(-18.4\pm0.3)$ & $0.98\pm0.01$ \\ \\
 (k) & $M_{\mbox{disk}}$-$M_{\mbox{bulge}}$ &  $M_{\mbox{bulge}}=(1.0\pm0.1)M_{\mbox{disk}}+(3.0\pm3.0)$ & $1.42\pm0.03$ \\
 & $|r|=0.69$, $p=4\cdot 10^{-10}$ &  $M_{\mbox{disk}}=(0.55\pm0.09)M_{\mbox{bulge}}+(-11.0\pm2.0)$ & $1.03\pm0.02$ \\ \\
 (l) & $\log r_e$-$M_{\mbox{bulge}}$ &  $M_{\mbox{bulge}}=(-4.1\pm0.6)\log r_e+(-19.6\pm0.1)$ & $1.41\pm0.02$ \\
 & $|r|=0.67$, $p=3\cdot 10^{-09}$ &  $\log r_e=(-0.14\pm0.02)M_{\mbox{bulge}}+(-2.9\pm0.3)$ & $0.262\pm0.004$ \\ \\
 (m) & $\log V_{\mbox{max}}$-$\log r_{50}$ &  $\log r_{50}=(1.1\pm0.2)\log V_{\mbox{max}}+(-1.8\pm0.4)$ & $0.204\pm0.007$ \\
 & $|r|=0.65$, $p=1\cdot 10^{-08}$ &  $\log V_{\mbox{max}}=(0.5\pm0.1)\log r_{50}+(1.95\pm0.06)$ & $0.134\pm0.005$ \\ \\
 (n) & $\log V_{\mbox{max}}$-$\log h$ &  $\log h=(1.0\pm0.1)\log V_{\mbox{max}}+(-1.7\pm0.3)$ & $0.205\pm0.008$ \\
 & $|r|=0.63$, $p=3\cdot 10^{-08}$ &  $\log V_{\mbox{max}}=(0.46\pm0.06)\log h+(1.98\pm0.04)$ & $0.141\pm0.004$ \\ \\
 (o) & $\log h$-$M_{\mbox{bulge}}$ &  $M_{\mbox{bulge}}=(-4.6\pm0.8)\log h+(-16.3\pm0.4)$ & $1.66\pm0.02$ \\
 & $|r|=0.60$, $p=3\cdot 10^{-07}$ &  $\log h=(-0.09\pm0.01)M_{\mbox{bulge}}+(-1.1\pm0.3)$ & $0.231\pm0.002$ \\ \\
 (p) & $\log r_{50}$-$M_{\mbox{bulge}}$ &  $M_{\mbox{bulge}}=(-3.7\pm0.8)\log r_{50}+(-16.6\pm0.5)$ & $1.83\pm0.02$ \\
 & $|r|=0.53$, $p=1\cdot 10^{-05}$ &  $\log r_{50}=(-0.08\pm0.02)M_{\mbox{bulge}}+(-0.8\pm0.4)$ & $0.260\pm0.004$ \\ \\
 (q) & $\log h$-$\log r_e$ &  $\log r_e=(0.8\pm0.3)\log h+(-0.6\pm0.2)$ & $0.41\pm0.01$ \\
 & $|r|=0.48$, $p=9\cdot 10^{-05}$ &  $\log h=(0.3\pm0.2)\log r_e+(0.62\pm0.03)$ & $0.28\pm0.01$ \\ \\
 (r) & $\log r_{50}$-$\log r_e$ &  $\log r_e=(0.6\pm0.2)\log r_{50}+(-0.5\pm0.1)$ & $0.362\pm0.002$ \\
 & $|r|=0.41$, $p=1\cdot 10^{-03}$ &  $\log r_{50}=(0.3\pm0.1)\log r_e+(0.68\pm0.02)$ & $0.275\pm0.003$ \\ \\
\hline 
\end{tabular} 
\end{table*}

All equations in table \ref{tab:corr} can be used as a way of obtaining
approximate physical parameters for one given measurement as well as for
constraining models of galaxy formation at the current cosmic time. Out of the 21
pairs of parameters, we observe that only three combinations have relatively low
correlation coefficients. This indicates that most of the spiral galaxy
properties are somehow linked. Although there are many possible properties that
shape galaxies, such as different mass, angular momentum, and despite secular
evolution effects, such as those which may form bars, there is still a great
similarity among spirals which is still to be explained. Also, this large number
of correlations restrict the interpretation of the correlations
individually, and certainly a comprehensible interpretation will be possible only
 with a more complete model of galaxy evolution
\citep[see][]{2010ApJ...720L..72S}. Nevertheless, we are going to briefly
discuss a few of the scaling laws that have been observed here and previously in the literature,
with the exception of the Tully-Fisher relation \citep{1977A&A....54..661T},
which we address with greater detail in \ref{sec:tf}.

The relation (q) between the sizes of bulges and disks was obtained previously
by other authors \citep{1996ApJ...457L..73C,2005A&A...434..109A} and may have
important clues for galaxy formation scenarios. \citet{1996ApJ...457L..73C}
argue that disks should have been formed earlier than bulges and, therefore, the
properties of the bulges are linked to their host disks. Due to the relatively
low S\'ersic indices of the bulges, these are indeed likely to be
pseudobulges \citep{2008AJ....136..773F}, which are formed by secular evolution
of the disks and, therefore, correlations among these parameters naturally arise
in a scenario of secular evolution, disfavoring scenarios of decoupled size
relation such as bulges formed by mergers. We have found that the median value
of $r_e/h$ is 0.14 considering all galaxies, which is in agreement to the values
in the literature \citep[for instance,][]{2010MNRAS.405.1089L}.

The luminosity of bulge is also of importance to understand its origin. Bulge 
luminosities and sizes are expected to correlate, as already indicated in
equation \eqref{eq:msersic},  $L_{\mbox{bulge}}\propto r_e^2$, and indeed there
is a strong correlation as shown in equation (l). Moreover, the bulge luminosity
is correlated to all other measured properties of the disks, to the whole galaxy
and also to the rotation velocity. Therefore, the properties of the bulges we
observe in late-type galaxies of the local universe are probably the result
of secular evolution. Interestingly, the  bulge luminosity is also correlated
with the supermassive black hole masses \citep{1995ARA&A..33..581K},
illustrating the important role of the bulges to understand the processes of
galaxy formation yet to be fully understood. 

Another parameter that correlates strongly with almost all the others is the luminosity of the
galaxy, as shown in equations (a), (b), (f), (g) and (j). The importance of
the total luminosity, also observed by \citet{2007ApJ...671..203C}, may indicate
that the baryonic portion of the galaxy has a pivotal role in the appearance of
galaxies: it is connected with the gravitational potential through the velocity,
but has a more direct link with the size (or shape) of the galaxy.  

Other photometric relations well documented in the literature include the
relation (k) between bulge and disk luminosities \citep{2010MNRAS.405.1089L},
and (f) which relates bulge and total luminosities  \citep{2007ApJ...658..960C}. 
The rotation velocity of the 
galaxies is usually studied in comparison with integrated photometry, such as
given in relation (b), the Tully-Fisher relation, and the size-velocity
relation (m) also studied by \citet{2007ApJ...671..203C}. However, here we show that
the rotation velocity also strongly correlate with the luminosity and size of
the disk component, as shown in relations (c) and (n), which is expected because
the disk is responsible for the majority of the light of the galaxy.
Interestingly, however, the luminosity of the bulge also correlates
with the rotation velocity, as shown in relation (h), indicating that the bulge
properties have a dynamical link with the galaxy that hosts it.

\subsection{Tully-Fisher relation} \label{sec:tf}

The Tully-Fisher relation \citep[hereafter TF relation,][]{1977A&A....54..661T}
is the most important scaling relation for disk galaxies, and it has been used
for several purposes including distance determinations and, historically, as a
way of measuring the Hubble constant. The TF relation relates the maximum velocity of the
rotation curve, $V_{\text{max}}$ with the total magnitude of the galaxies in the
form of a power law. The TF relation has already been measured in the \ac{ghasp}
sample previously by \citet{2011MNRAS.416.1936T} in the near-infrared bands H and K,
so here we add to those results the optical \Rc-band. We adopt the following
parametrization

\begin{equation} M_\lambda = \alpha_\lambda \cdot \log \left ( \frac{V_{\rm
max}}{\text{km/s}}\right ) + \beta_\lambda \text{,} \label{eq:tf} \end{equation}

\noindent where $M_\lambda$ is the absolute total magnitude in the passband
$\lambda$, $V_{\rm max}$ is the maximum velocity of the rotation curve,
$\alpha_\lambda$ and $\beta_\lambda$ are the slope and the zero point of the TF
relation. Both the slope and the zero point of the Tully Fisher relation are of
importance because they may be used either as constraints or as tests for models
of galaxy formation and evolution.

To produce a suitable sample for this specific relation, we select the galaxies
according to the following criteria. We remove galaxies with inclinations
greater than $75$\Deg due to their high internal extinction, and also galaxies
with inclinations smaller than $20$\Deg because of their higher uncertainty in
the determination of the rotation curve velocity. We also exclude galaxies with
recession velocities lower than 3000 km s$^{-1}$ due to possible peculiar
velocities affecting the Hubble flow, except for the cases where more accurate
distance indicators were used, such as Cepheids and red-giant branch distances.
Finally, as we are only dealing with H$\alpha$ velocity fields, we use the
analysis of \citet{2008MNRAS.388..500E} to exclude from the sample galaxies for
which the maximum rotation velocity is not achieved according to their
classification of the maps. We use our two absolute magnitude estimators, the
asymptotic ($M_{\rm R,total}$) and the isophotal ($M_{\rm R,23.5}$) as the probe
of the galaxy luminosity, resulting in samples with 80 and 72 galaxies
respectively. Most galaxies of \ac{ghasp} sample are not part of clusters of
galaxies, so we consider that our TF relation is basically probing the field
environment, although the expected difference of the TF relation in different
environments is mild \citep{2007ApJ...659.1172D,2012MNRAS.425..296M}. 

\begin{figure*} 
\centering 
\includegraphics[width=0.4\linewidth]{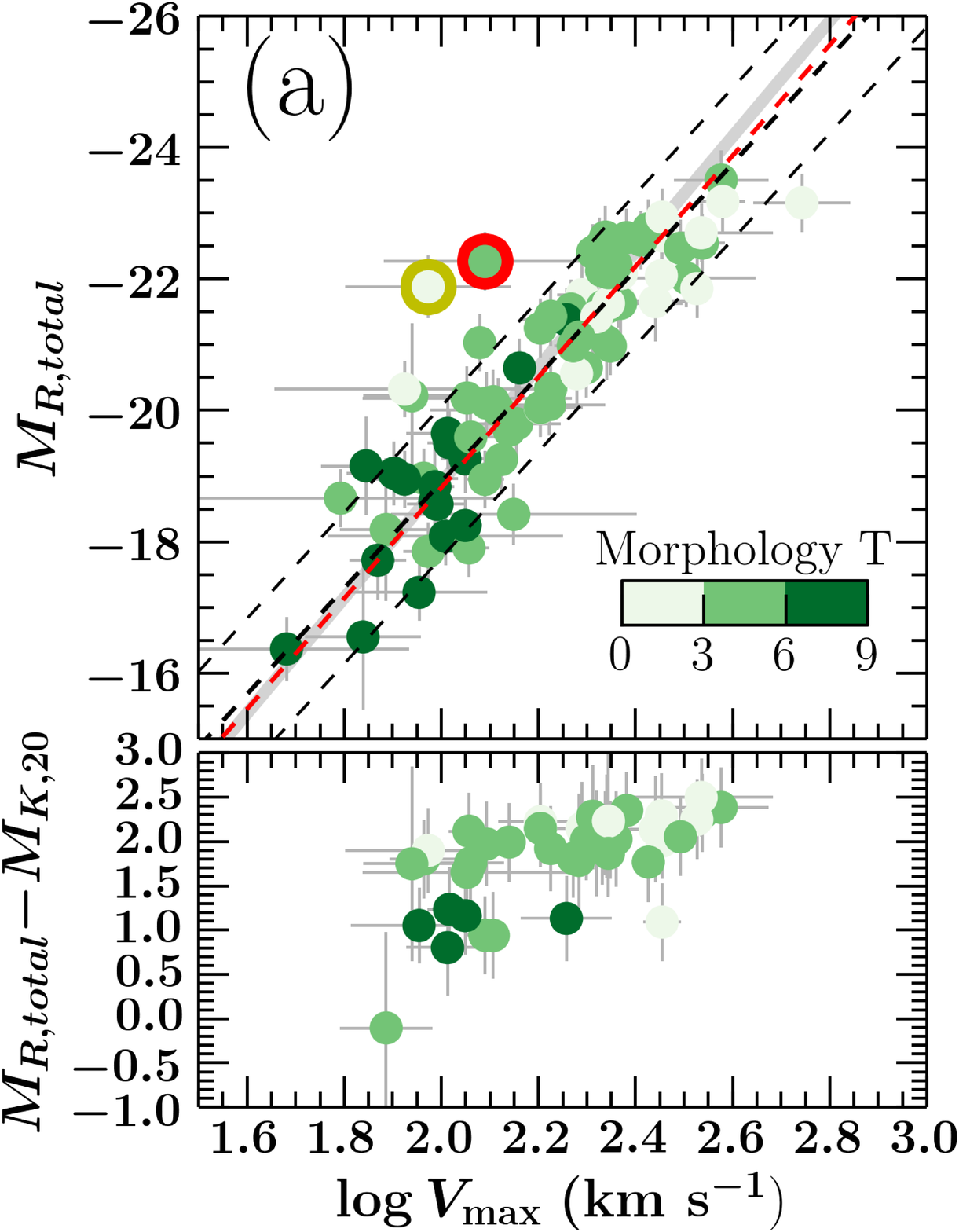}
\includegraphics[width=0.4\linewidth]{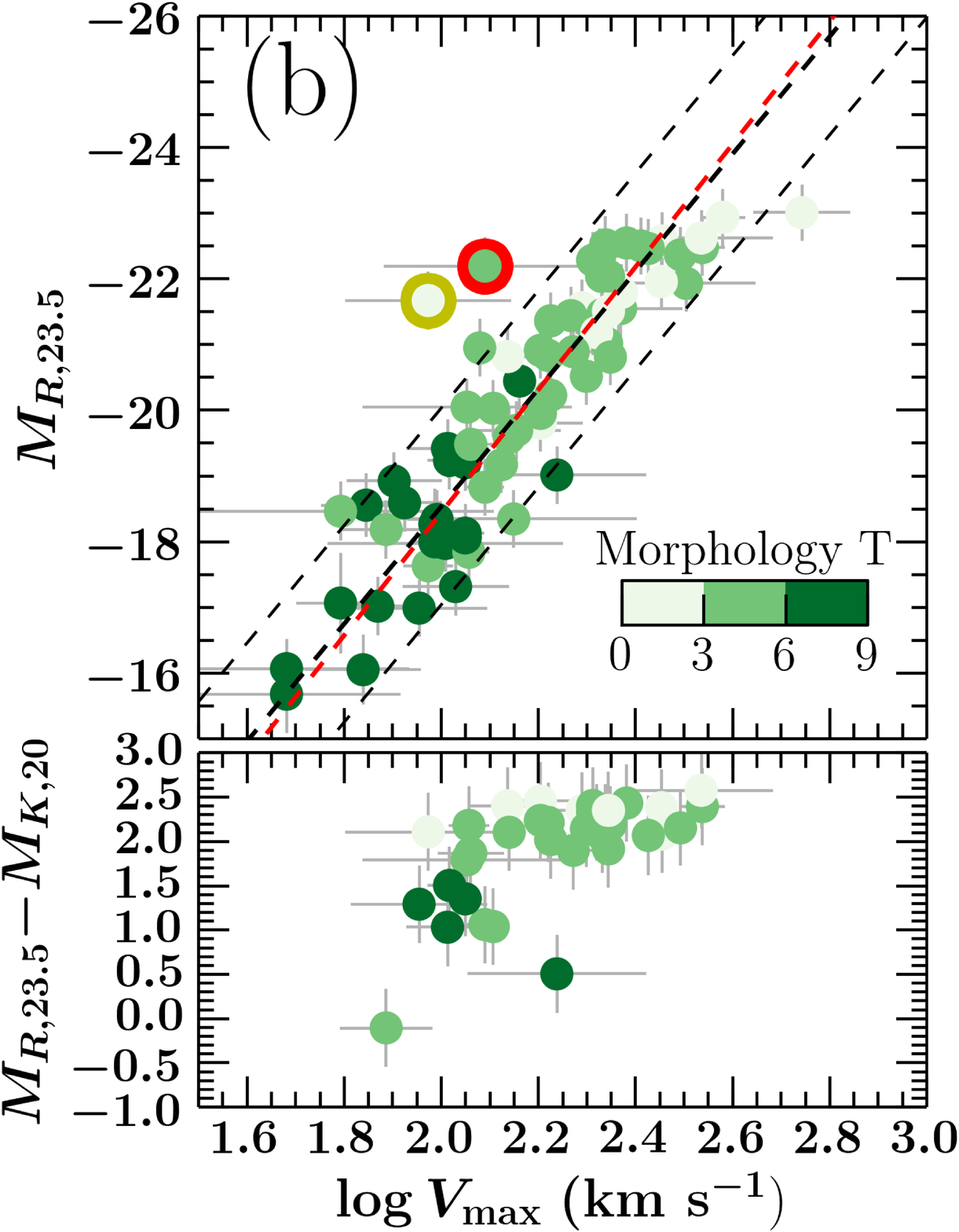} 
\caption{Tully-Fisher
relation for the \Rc-band for \ac{ghasp} galaxies considering (a) the total
asymptotic magnitude, $M_{\rm R,total}$, and (b) the isophotal total magnitude
inside the isophote of 23.5 mag arcsec$^{\rm -2}$. The gray line in panel (a)
indicates the results from \citet{2000ApJ...529..698S}. In the top panel of each
figure, the black dashed line indicates the direct TF relation results and the
best fit $\pm1\sigma_\beta$, while the red dashed line indicates the inverse TF
relation results. The bottom panels indicates the colour of the galaxies using the
K band total magnitudes used for the TF relation in \citet{2011MNRAS.416.1936T}
inside the isophote of 20 mag arcsec$^{\rm -2}$. The colour of each galaxy reflects 
the colourbar scale in the upper panels, separating objects according to their 
numerical morphology. The two objects highlighted with a halo are NGC
12276 (yellow) and NGC 4256 (red), which for different reasons are offset to 
the TF relation (see text for details).}
\label{fig:tullyfisherR} \end{figure*} 

The TF relation in the \Rc-band is shown in Figure \ref{fig:tullyfisherR}. To
calculate the regression coefficients, we have used the $\chi^2$ minimization of
equation \eqref{eq:chi2} as described in section \ref{subsec:corr}. Also, the
so-called inverse Tully-Fisher relation \citep{1980AJ.....85..801S} coefficients
are calculated as follows. We calculated the coefficients
$\alpha'$ and $\beta'$ by interchanging the variables $x_i\rightleftarrows y_i$
in equation \eqref{eq:linear}, and then calculated the inverse TF relation using
the relations $\alpha=1/\alpha'$ and $\beta=-\beta'/\alpha'$. The summary of the
results for the TF relation and the inverse TF relations is presented in Table
\ref{tab:tf}.

\begin{table} 
\centering 
\caption{Regression coefficients for the \Rc-band
Tully-Fisher relation using the total asymptotic magnitude and the total isophotal
magnitude including 80 and 72 galaxies respectively.} 
\label{tab:tf}
\begin{tabular}{lccc} \hline Relation &$\alpha_R$ & $\beta_R$ &
$\varepsilon_0$\\ \hline \multicolumn{4}{c}{Total asymptotic magnitude } \\ TF
&  $-8.1\pm0.5$  &  $-3.0\pm1.0$  &  $0.28\pm0.07$ \\ Inverse TF  &
$-8.4\pm0.7$  &  $-2.0\pm2.0$  &  $0.28\pm0.08$ \\ \hline
\multicolumn{4}{c}{Total isophotal magnitude } \\ TF  &  $-8.9\pm0.6$  &
$-1.0\pm1.0$  &  $0.33\pm0.06$ \\ Inverse TF  &  $-9.3\pm0.7$  &  $0.0\pm2.0$  &
$0.34\pm0.08$ \\ \hline \end{tabular} \end{table}

The TF relation is not just an important tool for measuring distances of
galaxies, but it is also crucial to highlight processes of galaxy evolution,
for instance, by comparing the TF relation of different morphological types.
Spiral galaxies have a single TF relation, but lenticular and elliptical
galaxies have TF relations which run approximately parallel when compared to spirals
\citep{2006MNRAS.373.1125B,2007ApJ...659.1172D}. We observe that the TF relation of 
spirals in the \Rc band is well defined for almost all galaxies, with only two
exceptions that are worth discussing. NGC 12276, marked in the figure 
\ref{fig:tullyfisherR} with a yellow halo, does not seem to have any 
special feature to be offset from the TF relation, so
one possibility to explain its position is that the distance to the galaxy is 
not accurate. We have used the value of 78.8 Mpc from Epinat et al. (2008) 
for consistency with the previous works, which is the expected value according 
to the   systematic velocity using the Hubble flow. However, 
\citet{1981ApJS...45..541P} have
estimated the distance to this galaxy of 40 Mpc using the ring size, which
implies in a difference of $\approx$1.5 magnitudes that is enough to bring the
galaxy much closer to the TF relation. The galaxy with a red halo in figure 
\ref{fig:tullyfisherR}, NGC 4256, has a peculiar morphology of a single arm and an
asymmetric rotation curve, which may be the cause to move the galaxy off 
the TF relation defined by relatively more relaxed spirals. In this case, star formation
may have been triggered recently as a response to a gravitational field, resulting 
in a relatively luminous object compared to the TF relation.

The slope and the zero point of the \Rc-band TF relation in our work are in
agreement with those previously derived the literature. \citet{2000ApJ...533..744T}, for instance, determined
the slope and zero point for a sample of 115 galaxies in four nearby clusters
with velocities derived from HI line widths, and by not considering errors in
both variables nor the intrinsic scatter of the relation, they have found
$\alpha_R=-7.65$ and $\beta_R=-4.3$, which is consistent with our results within 2 sigmas.
On the other hand, \citet{2001ApJ...563..694V} has derived the TF relation for
the Ursa Major cluster using the inverse relation without intrinsic scatter, and
fixing the uncertainties in 0.05 mag for the magnitudes and 5\% in the
velocities. In this framework, they found slopes ranging from -7.1 to -9, and
zero points ranging from -3.15 to 2.81 for their various samples, which is
similar to our inverse TF relation parameters. 

The TF relation in the infrared pass bands is important because these
wavelengths are reliable tracers of the stellar masses.
\citet{2011MNRAS.416.1936T} have used the \ac{ghasp} sample to derive the TF
relation in the H and K bands using 2MASS survey data
\citep{2006AJ....131.1163S} as well as stellar and baryonic TF relations, and
using a method similar to ours, they have obtained slopes of
$\alpha_H=-10.84\pm0.61$ and $\alpha_K=-11.07\pm0.63$ and zero points of
$\beta_H=1.97\pm1.36$ and $\beta_k=2.27\pm1.39$ for the H and K bands
respectively. As expected, the \Rc-band slope is greater than the slope in the
near-infrared band \citep[e.g.][]{2001ApJ...563..694V}. However, one important
feature observed in the infrared is a break in the TF relation for galaxies
with $\log V_{max} \lesssim 2.2$, in the sense that galaxies below this velocity
are under luminous related to the expected TF relation for bright galaxies. This
break in the TF relation is not noticed in the \Rc-band. This difference in the 
shapes of the near-infrared and optical TF at the low-mass regime can be understood 
if one inspects the bottom panels of Figure \ref{fig:tullyfisherR}, which show the 
optical to near-infrared colours of the galaxies as a function of the maximum rotation 
velocities. These panels show a flat colour distribution except for the galaxies with 
$V_{\rm max} \lesssim 125$ km s$^{\rm -1}$, indicating different mass-to-light 
ratios these galaxies. These low mass systems are bluer than more massive 
galaxies, indicating younger objects that may have been forming stars recently, 
and this effect fortuitously compensates for the difference in the stellar 
mass to light ratios, causing the differences in the shapes of the near-infrared and 
optical TFs at low masses. 

\section{Summary and Conclusion}

This study provided photometrically calibrated surface brightness profiles of
\ac{ghasp} galaxies, with 170 profiles in the \Rc-band and 108 in the \ac{sdss}
bands u,g,r,i and z. From these data, we derived \Rc-band integrated photometric
parameters, presented in Table \ref{table:photR}, which are consistent with other
works in the literature. All these results are public and will be available in
digital format at the Fabry Perot repository in
\url{http://cesam.lam.fr/fabryperot}.

We perform multi component structural decompositions in the \Rc-band, presented
in Table \ref{tab:decomp}, with the goal of separating the disk component from
bulges, bars, lenses and nuclear sources, as a preparation to our forthcoming
paper on the kinematic decomposition of \ac{ghasp} velocity fields, which will
be compared with the photometric work. 

Finally, we have applied new photometric data to observe bulges, disks and
global  scaling relations among luminosities, sizes and velocities in the
\Rc-band. We derived expressions for 18 scaling relations, which may be used to
constrain models of galaxy formation and evolution. In particular, we studied the
Tully Fisher relation using velocities derived solely from H$\alpha$ maps from
\ac{ghasp}. We have obtained slopes and zero points that are consistent with
previous findings in the literature in the \Rc-band for cluster galaxies,
reinforcing the idea that the Tully Fisher relation is basically a relation
between the total stellar content and the gravitational potential, which is
barely affected by the environment and the presence of photometric
substructures.

\section*{Acknowledgments}

We thank the \ac{ohp} technical team who has helped this project in several
ways, from support at the telescope to acquiring observational data. In
particular, we would like to thank Didier Gravalon, Jacky Taupenas and
Jean-Claude M\'evolhon who made several runs of observation. We also wish to
thank the several undergraduate students who had their first observing run
during the completion of this project under the supervision of DR and CA.  
We thank the anonymous referee for constructive comments which helped 
to improve this paper. CEB
and CMdO are grateful to FAPESP (Grants 2009/11236-0, 2011/21325-0 and
2006/56213-9) for financial support. PA, DR, BE, VP, CA and MM thank the PNCG
(Program National Cosmologie et Galaxies) for funding this project. CMdO and PA
thank USP-COFECUB for funding collaborative work between IAG and LAM. This
research has made use of the NASA/IPAC Extragalactic Database (NED) which is
operated by the Jet Propulsion Laboratory, California Institute of Technology,
under contract with the National Aeronautics and Space Administration. We
acknowledge the usage of the HyperLeda database
(\url{http://leda.univ-lyon1.fr}).

\begin{small} 

\end{small}


\begin{thebibliography}{99}

\bibitem[\protect\citeauthoryear{{Abazajian}, {Adelman-McCarthy}, {Ag{\"u}eros},
{Allam}, {Allende Prieto}, {An}, {Anderson}, {Anderson}, {Annis}, {Bahcall} and
et al.}{{Abazajian} et~al.}{2009}]{2009ApJS..182..543A} {Abazajian} K.~N.,
{Adelman-McCarthy} J.~K.,  {Ag{\"u}eros} M.~A.,  {Allam} S.~S.,  {Allende
Prieto} C.,  {An} D.,  {Anderson} K.~S.~J.,  {Anderson} S.~F., {Annis} J.,
{Bahcall} N.~A.,    et al., \apjs,  2009, vol.182, p. 543

\bibitem[Aguerri et al.(2005)]{2005A&A...434..109A} Aguerri, J.~A.~L.,
Elias-Rosa, N., Corsini, E.~M., \& Mu{\~n}oz-Tu{\~n}{\'o}n, C.\ 2005, \aap, 434,
109 

\bibitem[\protect\citeauthoryear{van Albada et al.}{1985}]{1985ApJ...295..305V}
van Albada T.~S., Bahcall J.~N., Begeman K., Sancisi R., 1985, ApJ, 295, 305 

\bibitem[van Albada 
\& Sancisi(1986)]{1986RSPTA.320..447V} van Albada, T.~S., \& Sancisi, R.\ 1986, Royal Society of London Philosophical Transactions Series A, 320, 447 

\bibitem[Amor{\'{\i}}n et al.(2007)]{2007A&A...467..541A} Amor{\'{\i}}n, R.~O.,
Mu{\~n}oz-Tu{\~n}{\'o}n, C., Aguerri, J.~A.~L., Cair{\'o}s, L.~M., \& Caon, N.\
2007, \aap, 467, 541 

\bibitem[\protect\citeauthoryear{Balcells et al.}{2003}]{2003ApJ...582L..79B}
Balcells M., Graham A.~W., Dom{\'{\i}}nguez-Palmero L., Peletier R.~F., 2003,
ApJ, 582, L79  

\bibitem[Bedregal et al.(2006)]{2006MNRAS.373.1125B} Bedregal, A.~G.,
Arag{\'o}n-Salamanca, A., \& Merrifield, M.~R.\ 2006, \mnras, 373, 1125 

\bibitem[\protect\citeauthoryear{Bernardi et al.}{2007}]{2007AJ....133.1741B}
{Bernardi}, M., {Hyde}, J.~B., {Sheth}, R.~K., {Miller}, C.~J., {Nichol}, R.~C.,
AJ, 133, 1741B

\bibitem[\protect\citeauthoryear{Bertin and Arnouts}{1996}]{1996A&AS..117..393B}
{Bertin}, E., {Arnouts}, S., \aas, 117, 393

\bibitem[Cabrera-Lavers \& Garz{\'o}n(2004)]{2004AJ....127.1386C}
Cabrera-Lavers, A., \& Garz{\'o}n, F.\ 2004, \aj, 127, 1386 

\bibitem[\protect\citeauthoryear{{Cair{\'o}s}, {Caon}, {V{\'{\i}}lchez},
{Gonz{\'a}lez-P{\'e}rez} and {Mu{\~n}oz-Tu{\~n}{\'o}n}}{{Cair{\'o}s}
et~al.}{2001}]{2001ApJS..136..393C} {Cair{\'o}s} L.~M.,  {Caon} N.,
{V{\'{\i}}lchez} J.~M., {Gonz{\'a}lez-P{\'e}rez} J.~N.,
{Mu{\~n}oz-Tu{\~n}{\'o}n} C., \apjs, 2001, vol.~136, p. 393

\bibitem[Carollo et al.(2007)]{2007ApJ...658..960C} Carollo, C.~M., Scarlata,
C., Stiavelli, M., Wyse, R.~F.~G., \& Mayer, L.\ 2007, \apj, 658, 960 

\bibitem[\protect\citeauthoryear{Chevalier \&
Ilovaisky}{1991}]{1991A&AS...90..225C} Chevalier C., Ilovaisky S.~A., 1991,
A\&AS, 90, 225 

\bibitem[Ciotti(1991)]{1991A&A...249...99C} Ciotti, L.\ 1991, \aap, 249, 99 

\bibitem[\protect\citeauthoryear{Courteau}{1996}]{1996ApJS..103..363C} Courteau
S., 1996, ApJS, 103, 363

\bibitem[Courteau et al.(1996)]{1996ApJ...457L..73C} Courteau, S., de Jong,
R.~S., \& Broeils, A.~H.\ 1996, \apjl, 457, L73 

\bibitem[Courteau et al.(2007)]{2007ApJ...671..203C} Courteau, S., Dutton,
A.~A., van den Bosch, F.~C., et al.\ 2007, \apj, 671, 203 

\bibitem[De Rijcke et al.(2007)]{007ApJ...659.1172D} De Rijcke, S., Zeilinger,
W.~W., Hau, G.~K.~T., Prugniel, P., \& Dejonghe, H.\ 2007, \apj, 659, 1172 

\bibitem[Driver et al.(2008)]{2008ApJ...678L.101D} Driver, S.~P., Popescu,
C.~C., Tuffs, R.~J., et al.\ 2008, \apjl, 678, L101 

\bibitem[Doyle et al.(2005)]{2005MNRAS.361...34D} Doyle, M.~T., Drinkwater,
M.~J., Rohde, D.~J., et al.\ 2005, \mnras, 361, 34 

\bibitem[\protect\citeauthoryear{Dutton et al.}{2005}]{2005ApJ...619..218D}
Dutton A.~A., Courteau S., de Jong R., Carignan C., 2005, ApJ, 619, 218

\bibitem[\protect\citeauthoryear{Epinat, Amram \& Marcelin}{2008}]
{2008MNRAS.390..466E} Epinat B., Amram P., Marcelin M., 2008, MNRAS, 390, 466E

\bibitem[\protect\citeauthoryear{Epinat et al.}{2008}]{2008MNRAS.388..500E}
Epinat B., Amram P., Marcelin M., Balkowski C., Daigle O., Hernandez O., Chemin
L., Carignan C., Gach J.-L., Balard P., 2008, MNRAS, 388, 500E

\bibitem[\protect\citeauthoryear{Epinat et al.}{2010}]{2010MNRAS.401.2113E}
Epinat B., Amram P., Balkowski C., Marcelin M., 2010, MNRAS, 401, 2113 

\bibitem[Erwin et al.(2005)]{2005ApJ...626L..81E} Erwin, P., Beckman, J.~E., \&
Pohlen, M.\ 2005, \apjl, 626, L81 

\bibitem[Erwin et al.(2008)]{2008AJ....135...20E} Erwin, P., Pohlen, M., \&
Beckman, J.~E.\ 2008, \aj, 135, 20  

\bibitem[\protect\citeauthoryear{{Fisher} and {Drory}}{{Fisher} and
{Drory}}{2008}]{2008AJ....136..773F} {Fisher} D.~B.,  {Drory} N., \aj, 2008,
vol.~136, p. 773
  
\bibitem[\protect\citeauthoryear{{Fisher} and {Drory}}{{Fisher} and
{Drory}}{2010}]{2010ApJ...716..942F} {Fisher} D.~B.,  {Drory} N., \apj, 2010,
vol.~716, p. 942

\bibitem[\protect\citeauthoryear{{Freeman}}{{Freeman}}{1970}]{1970ApJ...160..811F}
{Freeman} K.~C.,  \apj, 1970, vol.~160, p. 811

\bibitem[\protect\citeauthoryear{Gadotti}{2011}]{2011MNRAS.415.3308G} Gadotti
D.~A., 2011, MNRAS, 415, 3308 

\bibitem[\protect\citeauthoryear{Garrido et al.}{2002}]{2002A&A...387..821G}
Garrido O., Marcelin M., Amram P., Boulesteix J., 2002, A\&A, 387, 821 

\bibitem[\protect\citeauthoryear{Garrido et al.}{2003}]{2003A&A...399...51G}
Garrido O., Marcelin M., Amram P., Boissin O., 2003, A\&A, 399, 51 

\bibitem[\protect\citeauthoryear{Garrido, Marcelin, \&
Amram}{2004}]{2004MNRAS.349..225G} Garrido O., Marcelin M., Amram P., 2004,
MNRAS, 349, 225 

\bibitem[\protect\citeauthoryear{Garrido et al.}{2005}]{2005MNRAS.362..127G}
Garrido O., Marcelin M., Amram P., Balkowski C., Gach J.~L., Boulesteix J.,
2005, MNRAS, 362, 127

\bibitem[Gil de Paz et al.(2003)]{2003ApJS..147...29G} Gil de Paz, A., Madore,
B.~F., \& Pevunova, O.\ 2003, \apjs, 147, 29 

\bibitem[\protect\citeauthoryear{Haynes \&
Giovanelli}{1984}]{1984AJ.....89..758H} Haynes M.~P., Giovanelli R., 1984, AJ,
89, 758

\bibitem[Heraudeau \& Simien(1996)]{1996A&AS..118..111H} Heraudeau, P., \&
Simien, F.\ 1996, \aaps, 118, 111 

\bibitem[Hern{\'a}ndez-Toledo et al.(2007)]{2007AJ....134.2286H}
Hern{\'a}ndez-Toledo, H.~M., Zendejas-Dom{\'{\i}}nguez, J., \& Avila-Reese, V.\
2007, \aj, 134, 2286 

\bibitem[Hern{\'a}ndez-Toledo \& Ortega-Esbr{\'{\i}}(2008)]{2008A&A...487..485H}
Hern{\'a}ndez-Toledo, H.~M., \& Ortega-Esbr{\'{\i}}, S.\ 2008, \aap, 487, 485 

\bibitem[Hickson et al.(1989)]{1989ApJS...70..687H} Hickson, P., Kindl, E., \&
Auman, J.~R.\ 1989, \apjs, 70, 687 

\bibitem[Huchra(1977)]{1977ApJS...35..171H} Huchra, J.~P.\ 1977, \apjs, 35, 171 

\bibitem[\protect\citeauthoryear{van der Hulst, van Albada, \&
Sancisi}{2001}]{2001ASPC..240..451V} van der Hulst J.~M., van Albada T.~S.,
Sancisi R., 2001, ASPC, 240, 451 

\bibitem[James et al.(2004)]{2004A&A...414...23J} James, P.~A., Shane, N.~S.,
Beckman, J.~E., et al.\ 2004, \aap, 414, 23 

\bibitem[\protect\citeauthoryear{{Jedrzejewski}}{{Jedrzejewski}}{1987}]
{1987MNRAS.226..747J} {Jedrzejewski} R.~I., \mnras, 1987, vol.~226, p. 747

\bibitem[\protect\citeauthoryear{{Jester}, {Schneider}, {Richards}, {Green},
{Schmidt}, {Hall}, {Strauss}, {Vanden Berk}, {Stoughton}, {Gunn}, {Brinkmann},
{Kent}, {Smith}, {Tucker} and {Yanny}}{{Jester} et~al.}{2005}]
{2005AJ....130..873J} {Jester} S.,  {Schneider} D.~P.,  {Richards} G.~T.,
{Green} R.~F.,  {Schmidt} M.,  {Hall} P.~B.,  {Strauss} M.~A.,  {Vanden Berk}
D.~E.,  {Stoughton} C., {Gunn} J.~E.,  {Brinkmann} J.,  {Kent} S.~M.,  {Smith}
J.~A.,  {Tucker} D.~L.,    {Yanny} B., \aj, 2005, vol.~130, p. 873I c

\bibitem[\protect\citeauthoryear{{de Jong} and {van der Kruit}}{{de Jong} and
{van der Kruit}}{1994}]{1994A&AS..106..451D} {de Jong} R.~S.,  {van der Kruit}
P.~C., \aas, 1994, vol.~106, p. 451

\bibitem[\protect\citeauthoryear{Kassin, de Jong, \&
Pogge}{2006}]{2006ApJS..162...80K} Kassin S.~A., de Jong R.~S., Pogge R.~W.,
2006, ApJS, 162, 80 

\bibitem[\protect\citeauthoryear{Kassin, de Jong, \&
Weiner}{2006}]{2006ApJ...643..804K} Kassin S.~A., de Jong R.~S., Weiner B.~J.,
2006, ApJ, 643, 804 

\bibitem[Kent(1984)]{1984ApJS...56..105K} Kent, S.~M.\ 1984, \apjs, 56, 105 

\bibitem[\protect\citeauthoryear{Kent}{1986}]{1986AJ.....91.1301K} Kent S.~M.,
1986, AJ, 91, 1301 

\bibitem[Kormendy \& Richstone(1995)]{1995ARA&A..33..581K} Kormendy, J., \&
Richstone, D.\ 1995, \araa, 33, 581 

\bibitem[Kriwattanawong et al.(2011)]{2011A&A...527A.101K} Kriwattanawong, W.,
Moss, C., James, P.~A., \& Carter, D.\ 2011, \aap, 527, A101

\bibitem[van der Kruit \& Searle(1982)]{1982A&A...110...61V} van der Kruit,
P.~C., \& Searle, L.\ 1982, \aap, 110, 61

\bibitem[\protect\citeauthoryear{Landolt}{1992}]{1992AJ....104..340L} Landolt A.
U., 1992, AJ, 104, 340L 

\bibitem[\protect\citeauthoryear{Lauer et al.}{2007}]{2007ApJ...662..808L}
{Lauer}, T.~R., {Faber}, S.~M., {Richstone}, D., {Gebhardt}, K., {Tremaine}, S.,
{Postman}, M., {Dressler}, A., {Aller}, M.~C., {Filippenko}, A.~V., {Green}, R.,
{Ho}, L.~C., {Kormendy}, J., {Magorrian}, J., {Pinkney}, J., ApJ, 662, 808L

\bibitem[Laurikainen et al.(2010)]{2010MNRAS.405.1089L} Laurikainen, E., Salo,
H., Buta, R., Knapen, J.~H., \& Comer{\'o}n, S.\ 2010, \mnras, 405, 1089 

\bibitem[\protect\citeauthoryear{Lisker et al.}{2007}]{2007ApJ...660.1186L}
Lisker, T., Grebel, E.~K., {Binggeli}, B., {Glatt}, K., 2007, ApJ, 660, 1186L

\bibitem[\protect\citeauthoryear{{MacArthur}, {Courteau} and
{Holtzman}}{{MacArthur} et~al.}{2003}]{2003ApJ...582..689M} {MacArthur} L.~A.,
{Courteau} S.,    {Holtzman} J.~A., \apj, 2003, vol.~582, p. 689

\bibitem[\protect\citeauthoryear{Mart{\'{\i}}n-Navarro et
al.}{2012}]{2012MNRAS.427.1102M} Mart{\'{\i}}n-Navarro I., et al., 2012, MNRAS,
427, 110

\bibitem[Matthews \& Uson(2008)]{2008AJ....135..291M} Matthews, L.~D., \& Uson,
J.~M.\ 2008, \aj, 135, 291 

\bibitem[\protect\citeauthoryear{McDonald, Courteau, \&
Tully}{2009}]{2009MNRAS.394.2022M} McDonald M., Courteau S., Tully R.~B., 2009,
MNRAS, 394, 2022

\bibitem[Mocz et al.(2012)]{2012MNRAS.425..296M} Mocz, P., Green, A., Malacari,
M., \& Glazebrook, K.\ 2012, \mnras, 425, 296

\bibitem[\protect\citeauthoryear{Moffat}{1969}]{1969A&A.....3..455M} Moffat
A.~F.~J., 1969, A\&A, 3, 455

\bibitem[\protect\citeauthoryear{{Noordermeer} and {van der
Hulst}}{{Noordermeer} and {van der Hulst}}{2007}]{2007MNRAS.376.1480N}
{Noordermeer} E.,  {van der Hulst} J.~M., \mnras, 2007, vol.~376, p. 1480

\bibitem[Patterson(1940)]{1940BHarO.914....9P} Patterson, F.~S.\ 1940, Harvard
College Observatory Bulletin, 914, 9 

\bibitem[Pedreros \& Madore(1981)]{1981ApJS...45..541P} Pedreros, M., \& Madore,
B.~F.\ 1981, \apjs, 45, 541 

\bibitem[Pohlen \& Trujillo(2006)]{2006A&A...454..759P} Pohlen, M., \& Trujillo,
I.\ 2006, \aap, 454, 759 

\bibitem[Press et al.(1992)]{1992nrfa.book.....P} Press, W.~H., Teukolsky, 
S.~A., Vetterling, W.~T., 
\& Flannery, B.~P.\ 1992, Cambridge: University Press, |c1992, 2nd ed.,  

\bibitem[De Rijcke et al.(2007)]{2007ApJ...659.1172D} De Rijcke, S., 
Zeilinger, W.~W., Hau, G.~K.~T., Prugniel, P., 
\& Dejonghe, H.\ 2007, \apj, 659, 1172 

\bibitem[\protect\citeauthoryear{Rubin, Thonnard, \&
Ford}{1978}]{1978ApJ...225L.107R} Rubin V.~C., Thonnard N., Ford W.~K., Jr.,
1978, ApJ, 225, L107

\bibitem[Schechter(1980)]{1980AJ.....85..801S} Schechter, P.~L.\ 1980, \aj, 85,
801 

\bibitem[Sakai et al.(2000)]{2000ApJ...529..698S} Sakai, S., Mould, J.~R.,
Hughes, S.~M.~G., et al.\ 2000, \apj, 529, 698 

\bibitem[\protect\citeauthoryear{{Schlegel}, {Finkbeiner} and
{Davis}}{{Schlegel} et~al.}{1998}]{1998ApJ...500..525S} {Schlegel} D.~J.,
{Finkbeiner} D.~P.,    {Davis} M., \apj, 1998, vol.~500, p. 525

\bibitem[Sersic(1968)]{sersic68} Sersic, J.~L.\ 1968, Cordoba, 
Argentina: Observatorio Astronomico, 1968 

\bibitem[Shen et al.(2010)]{2010ApJ...720L..72S} Shen, J., Rich, R.~M.,
Kormendy, J., et al.\ 2010, \apjl, 720, L72 

\bibitem[Skrutskie et al.(2006)]{2006AJ....131.1163S} Skrutskie, M.~F., Cutri,
R.~M., Stiening, R., et al.\ 2006, \aj, 131, 1163 

\bibitem[\protect\citeauthoryear{Spano et al.}{2008}]{2008MNRAS.383..297S} Spano
M., Marcelin M., Amram P., Carignan C., Epinat B., Hernandez O., 2008, MNRAS,
383, 297 


\bibitem[\protect\citeauthoryear{Stoughton et al.}{2002}]{2002AJ....123..485S}
Stoughton C., et al., 2002, AJ, 123, 485 

\bibitem[Taylor et al.(2005)]{2005ApJ...630..784T} Taylor, V.~A., Jansen, R.~A.,
Windhorst, R.~A., Odewahn, S.~C., \& Hibbard, J.~E.\ 2005, \apj, 630, 784 

\bibitem[Thomas et al.(2008)]{2008A&A...486..755T} Thomas, C.~F., Moss, C.,
James, P.~A., et al.\ 2008, \aap, 486, 755 

\bibitem[\protect\citeauthoryear{Torres-Flores et
al.}{2011}]{2011MNRAS.416.1936T} Torres-Flores S., Epinat B., Amram P., Plana
H., Mendes de Oliveira C., 2011, MNRAS, 416, 1936

\bibitem[Tremaine et al.(2002)]{2002ApJ...574..740T} Tremaine, S., Gebhardt, K.,
Bender, R., et al.\ 2002, \apj, 574, 740 

\bibitem[Trujillo et al.(2001)]{2001MNRAS.328..977T} Trujillo, I., Aguerri,
J.~A.~L., Cepa, J., \& Guti{\'e}rrez, C.~M.\ 2001, \mnras, 328, 977 

\bibitem[Tully 
\& Fisher(1988)]{1988cng..book.....T} Tully, R.~B., \& Fisher, J.~R.\ 1988, Catalog of Nearby Galaxies, by R.~Brent Tully and J.~Richard Fisher, pp.~224.~ISBN 0521352991.~Cambridge, UK: Cambridge University Press, April 1988., 

\bibitem[\protect\citeauthoryear{{Tully} and {Fisher}}{{Tully} and
{Fisher}}{1977}]{1977A&A....54..661T} {Tully} R.~B.,  {Fisher} J.~R., \aaa,
1977, vol.~54, p. 661

\bibitem[Tully et al.(1996)]{1996AJ....112.2471T} Tully, R.~B., Verheijen,
M.~A.~W., Pierce, M.~J., Huang, J.-S., \& Wainscoat, R.~J.\ 1996, \aj, 112, 2471 

\bibitem[Tully et al.(1998)]{1998AJ....115.2264T} Tully, R.~B., Pierce, M.~J.,
Huang, J.-S., et al.\ 1998, \aj, 115, 2264 

\bibitem[\protect\citeauthoryear{{Tully} and {Pierce}}{{Tully} and
{Pierce}}{2000}]{2000ApJ...533..744T} {Tully} R.~B.,  {Pierce} M.~J., \apj,
2000, vol.~533, p. 744

\bibitem[\protect\citeauthoryear{de Vaucouleurs}{1948}]{1948AnAp...11..247D} de
Vaucouleurs G., 1948, AnAp, 11, 247 

\bibitem[de Vaucouleurs(1958)]{1958ApJ...128..465D} de Vaucouleurs, G.\ 1958,
\apj, 128, 465 

\bibitem[\protect\citeauthoryear{Verheijen}{2001}]{2001ApJ...563..694V}
Verheijen M.~A.~W., 2001, ApJ, 563, 694 

\bibitem[\protect\citeauthoryear{Zwicky}{1937}]{1937ApJ....86..217Z} Zwicky F.,
1937, ApJ, 86, 217 \end{thebibliography}
\end{document}